\documentclass[10pt]{article}
\usepackage[margin=1.0in,footskip=0.25in]{geometry}
\usepackage{authblk}
\usepackage{graphicx}
\usepackage{subfig}
\usepackage{listings,xcolor}
\usepackage{amsmath}
\usepackage{amsfonts}
\usepackage{amssymb}
\usepackage[numbers]{natbib}
\usepackage{booktabs}  
\usepackage{tabularx}
\usepackage{tikz}
\usetikzlibrary{positioning}
\usepackage{multirow}
\usepackage[title]{appendix}
\usepackage[plain]{fancyref}  
\usepackage[]{hyperref}
\usepackage{array} 
\usepackage{float} 
\newcolumntype{Y}{>{\centering\arraybackslash}X}

\hypersetup{
    colorlinks=true,
    linkcolor=blue,
    filecolor=magenta,      
    urlcolor=cyan,
    pdftitle={Overleaf Example},
    pdfpagemode=FullScreen,
    }   
\lstdefinestyle{perl}
{
   language=Perl,
   basicstyle=\ttfamily,
   keywordstyle=\bfseries\color{green!40!black},
   commentstyle=\color{purple!40!black},
   identifierstyle=\color{blue},
   stringstyle=\color{orange},
   alsoletter={\%},
}
\newcommand*{\fancyreflstlabelprefix}{lst}
\fancyrefaddcaptions{english}{%
  \providecommand*{\freflstname}{listing}%
  \providecommand*{\Freflstname}{Listing}%
}
\frefformat{plain}{\fancyreflstlabelprefix}{\freflstname\fancyrefdefaultspacing#1}
\Frefformat{plain}{\fancyreflstlabelprefix}{\Freflstname\fancyrefdefaultspacing#1}

\frefformat{vario}{\fancyreflstlabelprefix}{%
  \freflstname\fancyrefdefaultspacing#1#3%
}
\Frefformat{vario}{\fancyreflstlabelprefix}{%
  \Freflstname\fancyrefdefaultspacing#1#3%
}

\newcommand*{\fancyrefapndxlabelprefix}{apndx}
\fancyrefaddcaptions{english}{%
  \providecommand*{\frefapndxname}{appendix}%
  \providecommand*{\Frefapndxname}{Appendix}%
}
\frefformat{plain}{\fancyrefapndxlabelprefix}{\frefapndxname\fancyrefdefaultspacing#1}
\Frefformat{plain}{\fancyrefapndxlabelprefix}{\Frefapndxname\fancyrefdefaultspacing#1}

\frefformat{vario}{\fancyrefapndxlabelprefix}{%
  \frefapndxname\fancyrefdefaultspacing#1#3%
}
\Frefformat{vario}{\fancyrefapndxlabelprefix}{%
  \Frefapndxname\fancyrefdefaultspacing#1#3%
}

\newcommand\C[1]\null 
\DeclareUnicodeCharacter{202F}{FIX ME!!!!}  
\DeclareUnicodeCharacter{2009}{}  

\title{Enhancing non-Perl bioinformatic applications with Perl:\\[1ex] \large Building novel, component based applications using Object Orientation, PDL, Alien, FFI, Inline and OpenMP}
\author{Christos Argyropoulos MD, MSc, PhD, FASN}
\affil{Department of Internal Medicine, School of Medicine, University of New Mexico, Albuquerque, New Mexico, USA}
\date{\today}

\begin{document}
\maketitle

\begin{abstract}
Component-Based Software Engineering (CBSE) is a methodology that assembles pre-existing, re-usable software components into new applications, which is particularly relevant for fast moving, data-intensive fields such as bioinformatics. While Perl was used extensively in this field until a decade ago,more recent applications opt for a Bioconductor/R or Python. This trend represents a significantly missed opportunity for the rapid generation of novel bioinformatic applications out of pre-existing components since Perl offers a variety of abstractions that can facilitate composition.  In this paper, we illustrate the utility of Perl,for CBSE  through a combination of Object Oriented (OO) frameworks, the Perl Data Language (\texttt{PDL}) and facilities for interfacing with non-Perl code through Foreign Function Interfaces (FFI) and inlining of foreign source code. To do so, we enhance Polyester, a RNA sequencing simulator written in R, and edlib a fast sequence similarity search library based on text edit distance. The first case study illustrates the near effortless authoring of new, highly performant Perl modules for the simulation of random numbers using the GNU Scientific Library (GSL) and \texttt{PDL}, and proposes Perl and Perl/C alternatives to the Python tool cutadapt that is used to "trim" polyA tails from biological sequences. For the edlib case, we leverage the power of metaclass programming to endow edlib with coarse, process based parallelism, through the Many Core Engine (MCE) module and fine grained parallelism through OpenMP, a C/C++/Fortran Application Programming Interface (API) for shared memory multithreaded processing. These use cases provide proof-of-concept for the \texttt{Bio::SeqAlignment} framework, which can organize heterogeneous components in complex memory and command-line based workflows for the construction of novel bionformatic tools to analyze data from the long-read sequencing, e.g. Nanopore, sequencing platforms. 
\end{abstract}

\section{Introduction}
Component-Based Software Engineering (CBSE), assembles pre-existing, reusable software components into new applications. CBSE for scientific applications (SCBSE) has the potential to instill novel functionalities in existing, extensively tested components with the promise of rapid development cycles that are particularly relevant for fast moving, data-intensive fields such as bioinformatics. Dynamically typed, scripting, languages have played a particularly crucial role in SCBSE by acting as glue that integrates and coordinates hetero-geneous components or by facilitating communication of components that are organized as filters in complex data flows. These languages confer additional benefits such as rapid prototyping and support for multiple programming paradigms, that in turn allow effortless exploration of multiple architectural alternatives before settling to a final design. 

\section{Perl in Bioinformatics}
Perl was traditionally the go-to dynamically typed, scripting language in bioinformatics (“BioPERL”) due to its robust text manipulation capabilities, which was credited with saving the Human Genome Project nearly 30 years ago \cite{stein_how_1996}. Subsequently, Perl provided a platform independent way to construct \textit{pipeline frameworks}, i.e. a series of transformations between the output and input of (usually) command line tools organised in filters\cite{leipzig_review_2017}. With the winding down of the Human Genome Project in the early 2000s, the needs of bioinformaticians shifted partially away from sequencing to gene quantification and experimental analysis using statistical tools. This shift in needs provided an opportunity for languages that were deemed more appropriate for data analytics (e.g. Bioconductor/R initially and later Python) to enter this space. These structural changes combined with the reputation of Perl as a read-only language and the language's overall declining overall popularity has led to a diminishing creation of new Perl modules targeting bionformatics. The under-utilization of Perl in bioinformatics represents a significant missed opportunity to enhance applications in the field by leveraging: 
\begin{enumerate}
    \item the vast resources of Comprehensive Perl Archive Network (CPAN) featuring  
    \begin{itemize}
    \item a versatile and rich choice of Object Oriented (OO) modules that facilitate composition, while promoting creative component reuse.  
    \item an extensive, robust framework (\texttt{Alien}\footnote{\url{https://metacpan.org/pod/Alien}}) to fight the nemesis of dependency handling when building Perl applications that utilize \textit{external/foreign} libraries or tools, 
    \item multiple mature frameworks for interfacing with foreign libraries. Such applications can also leverage Perl’s rigorous framework for interacting with libraries in non-Perl languages using Foreign Function Interfaces (FFI, e.g. \texttt{Platypus::FFI}\footnote{\url{https://metacpan.org/pod/FFI::Platypus}}  for dynamic linking and the \texttt{Inline}\footnote{\url{https://metacpan.org/dist/Inline/view/lib/Inline.pod}} modules for both dynamic and static linking. These frameworks provide  in-script, alternatives to the more conventional bridges\cite{jenness_extending_2002} between Perl and C/C++ (XS, eXtendable Subroutine, SWIG, Simplified Wrapper and Interface Generator) that often require including additional, or highly technical in the case of XS,  interface files in one's modules. 
    \item the Perl Data Language (\texttt{PDL}\footnote{\url{https://metacpan.org/pod/PDL}}), a performant, auto-parallelizable, vectorized array extension that adds capabilities for multidimensional array handling, visualization, advanced numerical and statistical calculations to Perl\cite{glazebrook_pdl_1997}
\end{itemize}
    \item a shift away from traditional command-line, filter (in the Unix sense) communication based schemes for components of complex data flows. While the command line still reigns supreme in this area\cite{camacho_blast_2009,li_minimap2_2018,langmead_fast_2012}, many novel algorithms \cite{reinert_seqan_2017, daily_parasail_2016, sosic_edlib_2017} are delivered as both command-line tools AND libraries suitable for embedding into other applications. This feature offers a unique opportunity for memory-based alternatives to the traditional filter-based communication scheme that has been the mainstream approach in bioinformatics\cite{leipzig_review_2017}. 
    \item the increasing realization that complex,  production bioinformatic pipelines will benefit from evolving towards abstract class-based APIs for defining task dependencies that execute on multicore hardwre platforms or the cloud. 
\end{enumerate}
Perl's support for multiple computing paradigms, including procedural, functional, object oriented and even distributed, process based, programming seems ideally suited for this new world for bioinformatic applications. As previously pointed out, class-based approaches will be needed to achieve "a high level of granularity in terms of the concurrent processing of data"\cite{leipzig_review_2017}. It is precisely this insight, and the context of the author's applied work (\Fref{sec:RNADNAsequencing}) that provided the impetus of the Perl based object oriented and metaclass programming approach for bioinformatic tasks (\Fref{sec:PerlIn21stCenturySequencing}) described in this report.

\section{Problem Description and Requirements}
In this section we will provide a general, overview of the real world, applied problem of analysing data from long-read sequencing\cite{mackenzie_introduction_2023}, followed by the Perl solutions that were developed to address a few of the challenges we encountered during the introduction of a new protocol for RNA sequencing (PALS-NS)\cite{mackenzie_make_2022}.To decouple the presentation of the Perl solutions from the specifics of the PALS-NS sequencing workflow, we will utilize two case studies that highlight a newfound relevance of the language for rapid composition of solutions for what is currently the largest "market" in Bioinformatics, namely sequencing applications. 

\subsection{Of DNA and RNA Sequencing}\label{sec:RNADNAsequencing} 
Next Generation Sequencing (NGS) is an experimental methodology for the characterization of the genetic information (DNA sequencing) of an individual (human, animal, plant, or even cell). These techniques which evolved\cite{genomics_brief_2021} from manual identification of "ladder patterns" and radioactive detection on gel electrophoresis in the 1970s and 1980s to semi-automatic methods for the analyses of distinct color patterns in capillary electrophoresis devices (mid 1980s to 1990s), followed by pattern recognition of images of biochemical reactions on the surface of two dimensional "flow cells" (2000s to today). These techniques have continued to evolve since the first report of the sequencing of the genomes of Craig Venter, the leader of the private group that competed in the original Human Genome product, followed a few years later by the DNA pioneer James Watson\cite{wadman_james_2008}. While the Human Genome Project required 100 million dollars and a multi-national effort over nearly a decade, Professor Watson's sequencing ony took the effort of a handful scientists for four months and a price tag of 1.5 million dollars in 2008; fast forward in time, it is expected that the \$100 genome will be be widely available to everyone before the end of the present decade. Sequencing runs generate an enormous amount of data, e.g. obtaining one's one genome at the industry standard of x30-50 coverage (meaning each base, or "letter" in the DNA sequence is read 30-40 times) amounts to the equivalent of 100-150 GB of \textit{processed text} per person; the actual amount of raw data is closer to 200GB, necessitating computationally efficient workflows for processing and interpretation.
\\
Over the course of the last 2 decades the scope of sequencing expanded from genomic (DNA) to \textit{transcriptomic} (RNA) sequencing, allowing one to peek beyond the "storage" of genetic information to  the "trascription" of said information and the regulation of multiple genetic and epigenetic phenomena through RNA molecular intermediates. RNA sequencing, which often proceeds biochemically through \textit{in-vitro} DNA biochemistry comes with its own set of unique challenges for the bioinformatic analyst:
\begin{enumerate}
\item each RNA molecule will be present in variable amounts in the original sample, and this unequal representation creates issue for the identification of rare RNA species in the context of sequencing errors (equivalent to text corruption). In constrast, this unequal coverage is not present when sequencing DNA, considerably simplifying the ue of information for biological inferences. 
\item while the linear structure of the genome is more or less constant, RNA species are often scrambled (e.g. rearranged, "cut \& pasted" versions of their genomic counterparts. Hence, even the choice of the database used during searching can have a a large influence on the final results.
\item statistical techniques for identifying the evolutionary relationship between two DNA sequences are not well adapted to the problem of "similarity searching", a form of approximate text matching, between a given RNA sequence and the potential universe of RNA molecules it could have come from. This in turn requires the researcher to be open to explore algorithmic alternatives to traditional similarity search methods. In such a fluid environment, the need for new, more accurate and more performant software will almost never be satisfied. 
\item each of the previous steps have to be carried out independently for the tens of millions of sequences generated by a single RNA sequencing run. However, approximate text matching/similarity searching is a computationally demanding technique; the gold standard of sequence(text) alignment between a \textit{query} of length $m$ letters and a \textit{reference} sequence of length $n$ in a database exhibits worst case space and time complexity that scales as $O(mn)$. A large array of heuristic methods to execute such searches have been proposed over the last four decades in the literature\cite{song_new_2014,sahlin_effective_2021,bucher_sequence_1996,mitrophanov_statistical_2006}, with no clear path to make a choice between methods that are slow but reliable, and those that are fast by cutting corners. More importantly, an analyst that would like to propose and evaluate new (or hybrid) methods for similarity searching will often lack a versatile framework to compose new software for these analyses out of pre-existing components. 
\end{enumerate}
\subsubsection{Third Generation Sequencing Methods}
\begin{figure}
\centering  
\includegraphics[scale=0.65]{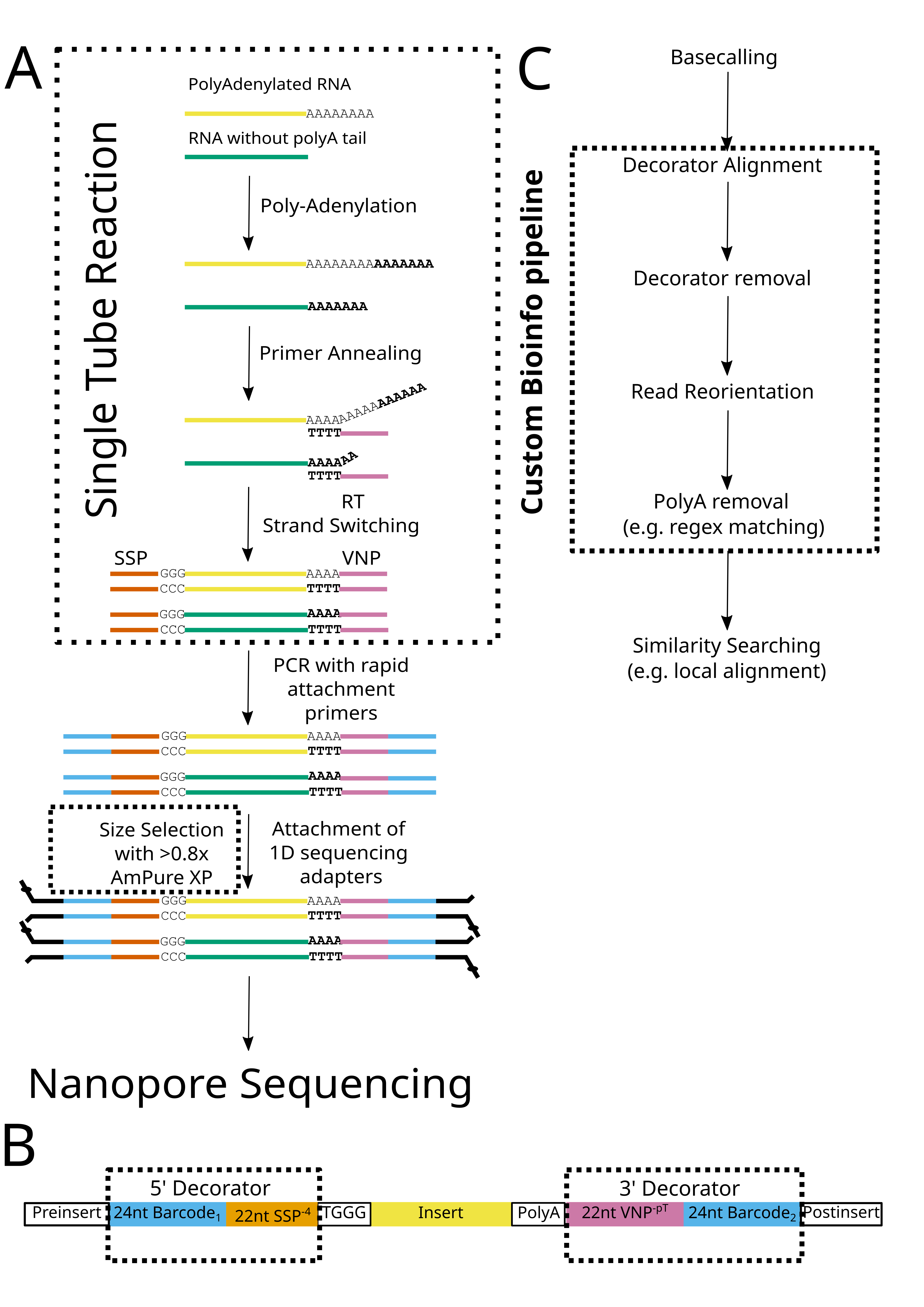}
\caption{PALS-NS experimental workflow (A), text- based model of a well-formed read (B) and custom bioinformatics pipeline (C) to remove decorators. Dashed boxes indicate modifications to biochemical protocols, read models and bioinformatics pipeline for the SMART protocol.
Abbreviations: SSP, Strand Switching Primer, PCR, Polymerase Chain Reaction, RT, Reverse Transcription, VNP, poly-thymidine based primer. Figure reproduced from \cite{mackenzie_make_2022}, available under the CC-BY 4.0 International license.}\label{fig:PALS-NS}
\end{figure}

Third generation sequencing approaches were introduced at the end of the 2000s \cite{athanasopoulou_third-generation_2021,van_dijk_third_2018} and provided a significant amount of innovation over the 2nd generation ones: the ability to generate longer reads (making it potentially easier to identify a given nucleic acid sequence) using a variety of a single molecule detection methodologies in real time, leading to lower cost, smaller, even portable devices (e.g. Oxford Nanopore Technologies Minion and Flongle platforms). In  contrast to early generation methods, the length of the sequences from long read platforms is not uniform complicating the  filtering of sequence features introduced by the biochemical reactions of the sequencing protocols. Furthermore, open platforms such as Nanopore sequencing don't confine the developer into a single biochemical workflow; in fact, one may devise nearly countless protocols that target detection of specific RNA modules, modifying the number and nature of the steps until the sample is loaded on the sequencer device. Invariably these modifications will end up introducting additional features to the sequences, i.e. additional text that will have to be filtered out prior to analysis. And here lies a problem: while computational pipelines for 2nd generation NGS could rely on positional information for such filtering (e.g. trim the first X and the last Y letters in a given sequence), long-read methods have to rely to some extent to approximate text matching/similarity searching to both filter out artificial sequence features AND identify (or map) the filtered text to the genome or transcriptome (the collection of RNA species in a given organism).
To make some of the issues more concrete let us consider the computational requirements for the analysis of sequencing data from the PALS-NS protocol\cite{mackenzie_make_2022}, a  modification of the original SMART RNA sequencing protocol\cite{zhu_reverse_2001} that was patented by the author for the simultaneous profiling of the entire universe of RNAs present in a cell. The basic idea of PALS-NS (\Fref{fig:PALS-NS}) is to artificially tag \textbf{all} RNA molecules in a given sample by attaching a long stretch of adenine bases \textit{polyA tail}, thus extending the scope of SMART from messenger RNA molecules (that naturally harbor such tails) to all RNA species; the polyA tail is an indispensable component of the biochemistry of the protocol, which allows capture and amplification of RNAs by the polymerase reaction. Each step in the biochemical wet lab workflow modifies the sequence present in the original sample ("inserts" in \Fref{fig:PALS-NS}) by introducing various artificial features. These extraneous features such as the polyA tail, as well as the "adapters" used to amplify sequences during PCR are nuisance "decorators", which have to be computationally removed from the text that is generated by the sequencers before a given insert can be mapped. Removal of the decorators is particularly important for short (20-30 letters long), regulatory RNAs, e.g. microRNAs that are 7-10 times shorter than the decorators and the polyA tails used to "fish" them out from the biological sample. These shorter RNAs are highly relevant for the development of new clinical diagnostics, for understanding complex biology and for the development of RNA interfering therapies. Hence, getting their identification just right is crucial for a wide spectrum of applications. 
\linebreak The original computational pipeline to remove the decorators in sequences generated by PALS-NS was written in Python as a collection of scripts that hardwired particular choices (e.g. the use of the software BLAST\cite{camacho_blast_2009} for similarity searching. While these choices deemed justified at the time, a closer re-examination of the problem needs reveal significant shortcomings in some of the initial component choices, which unfortunately had locked us in the design of the pipeline.  A subsequent (non-public) rewrite in Perl suggested substantial potential for flexible parameterization of the programming interface for nearly all steps in \Fref{fig:PALS-NS}C. Such parameterization could avoid hardwiring the algorithm for similarity searching for decorators, or the technique used to remove/trim the polyA tails, facilitating the exploration of multiple potential alternatives. 
\textbf{In summary, the increasing popularity and open nature of the 3rd generation sequencing methods, create an everlasting need for versatile, customizable and parallelizable software for the analysis of data that are generated by these methods.}

\subsection{Perl for SCBSE in Bioinformatics for the 21st century}\label{sec:PerlIn21stCenturySequencing}
The preceding exposition provides the general background of the challenges faced by anyone introducing a novel biochemical analysis workflow that requires the development of a more or less custom computational data flow and pipeline. The relevant issues are worth summarizing in a concise form in order to understand the rationale of the use cases we selected to work on:
\begin{description}
\item [Write the Software that Writes the Tests.] One needs a way to generate test data for the algorithmic innovations (or compositions!) one would like to throw at the computational workflow such as the one shown in \Fref{fig:PALS-NS}C. While actual bench work following the steps outlined in \Fref{fig:PALS-NS}A are crucial, real world experiments take money and effort to generate and analyse. Often one would like to test an algorithm against some known ground truth, and the easiest way to generate such data is through simulations. Sequencing simulators have been indispensable in evaluating new algorithms in the field, but they tend to target mature experimental workflows. For novel workflows, one will often need to modify an existing simulator, rather than write it from scratch, leading to the first case study we consider below, i.e. the modification of the Polyester RNA sequencing simulator.
\item [Fool Around to Find Out.] One needs a framework for rapidly combining components that may had not been tried out before to do something great (e.g. to provide an accurate, performant pipeline for the tasks shown in \Fref{fig:PALS-NS}C). At a minimum one should be free to step out of the traditional Unix filter approach for making components communicate. The amount of DRAM available in our computers means that one SHOULD do as many tasks as possible in memory, rather than relying on disk IO. At the same time, the number of processor cores available in compute platforms from cell phones to servers and from cell phones to GPU accelerators increases year by year, so why not farm out tasks to them? In the second case study, we thus take an existing, highly performant implementation for similarity searching and show how one can go beyond its command line version, to construct parallel sequence mappers. Both the extent and the granularity (process vs thread) of parallelism are tightly controlled by the user through the Perl programmer interface. 
\item [Abstract and  Parallelize.] We propose (and implemented) abstract classes for the two data flows that cover the vast majority (if not all) of sequence mapping scenarios scenarios. These scenarios illustrate the tremendous potential of Perl for abstraction of parallel applications using the \texttt{Task::Kensho} modules for Enlightened Perl development.
\item [Work with NUMA.] Non Uniform Memory Access (NUMA) machines, are a significant hardware player in bioinformatics. In fact, for the memory bound sequence similarity algorithms, the NUMA architecture may exert a considerable influence on performance\cite{lopez-villellas_genarchbench_2024}.  Therefore, our benchmarking was undertaken in the most accessible NUMA mchine, a dual socket Xeon workstation (see Appendix for the technical specifications) and certain benchmarking results were directly related to the NUMA topology of our target hardware.
\end{description}
A unifying theme of the two case studies is the  use of Perl's ligher object oriented modules to raise the level of abstraction during component composition. For the more challenging second case study we leveraged Perl's meta-programming frameworks in order to properly and rigorously abstract the problem of creating a sequence mapping application. The two case studies are conceptually linked through the polyA tagging of biological sequences and the need to remove them. We thus opted to highlight a couple of solutions for this task under the first case study. 

\subsection{Case Study 1: \texttt{Polyester}}
\texttt{Polyester}\cite{frazee_polyester_2015} is an R based RNA sequencing simulation of key steps in a RNA-seq experiment, including fragmentation, reverse-complementing, and sequencing, and produces files of RNA reads out of a set of reference sequences (usually in the FASTA flat text format). Polyester can simulate differential expression of similar versions of the same RNA ("isoforms") for a variety of experimental designs at the read level. This feature allows one to use the output of Polyester to test methods for mapping and RNA quantification of rather similar RNAs. One of the main \textbf{technical advantages} of Polyester over more recent simulators is the ability to simulate smaller RNA sequencing (RNA-seq) datasets, reducing the computing resources and time required to generate such datasets. To do so, \texttt{Polyester} does not provide methods to attach polyA tails (or the decorating features of SMART protocols), rather focusing on generating abbreviated and error-prone of the sequences of interest (inserts in \Fref{fig:PALS-NS}B). A straightforward \textit{enhacement} of \texttt{Polyester} to enable the simulation of sequences with polyA tails would attach those to the reference sequences \textit{before} those were provided to the software. To do so from within Perl one would need to implement (or get from CPAN) modules that can:
\begin{enumerate}
\item read, manipulate and write FASTA files in a performant manner. While the \texttt{BioPerl} and \texttt{FAST:BIO} distributions provide modules for reading such files, these implementations are geared towards the highly annotated reference sequences biological databases, instead of the featureless FASTA files that Polyester needs.  The \texttt{BioX:Seq} modules introduced in the early 2020s are not only more light, but they target FASTA files generated by sequencers and were thus chosen to provide the relevant IO functionalities. 
\item simulate random lengths of polyA tails using a variety of truncated statistical distributions. Truncated distributions restrict the random lengths to lie within certain windows, as one would expect from biochemical and biological processes. There are currently no modules in CPAN that can simulate truncated distributions; even R with its extensive statistical computing capabilities does not offer many choices for this task. Hence the simulation of such random variables had to be implemented from scratch in an extensible manner (and the first version of a new PDL module!).
\item a way to keep track of the modifications (length of the RNA tail) introduced by the polyA tail simulation step, so that algorithms for the trimming of the polyA tails could be evaluated against the ground truth.   
\end{enumerate}

\subsection{Case Study 2: \texttt{Edlib}}
\texttt{Edlib}\cite{sosic_edlib_2017} is an extremely fast lightweight C++ library for the alignment of two sequences and the assessment of their similarity using the Levenshtein (edit) distance\cite{berger_levenshtein_2021}. The edit distance is he minimum number of single-character edits (insertions, deletions or substitutions) required to change one sequence into the other; two identical sequences will have an edit distance of zero, with an increasing distance indicating increasing dissimilarity. The edit distance is a metric over strings, that has applications in any area in which one requires \textit{approximate string matching}, e.g. natural language processing, bioinformatics or even optical character recognition. One of the fastest algorithms for implementing edit distance searching is that by Myers\cite{myers_fast_1999} that improved over the search implemented by Unix's \textit{agrep} utility. \texttt{Edlib} improved upon the Myers algorithm.  The codebase available at the github repository provides build files that generate static and dynamic libraries, as well as a command line tool that can support a limited form of database search: that of searching a large number of queries against a \textit{single target}. To allow edlib in all its incarnations to be used in production, we:
\begin{enumerate}
\item abstracted the data flows during sequence mapping applications, treating the software component performing the sequence search as a black box. 
\item overcome the limited ability of the \texttt{Edlib} command-line aligner to carry out database searches and highlight the ability of the abstract class design to operate with command-line tools and heavy disk IO and with libraries exchanging data across languages.
\item used our abstractions to embed \texttt{Edlib} into a Perl application that exhibits user controlled levels of coarse and fine grained parallelism. 
\end{enumerate}
\newpage

\section{Results and Discussion}
\subsection{Enhancing Polyester}
The main module created for this use case provides an application, command-line tool that wraps over the \texttt{Polyester} simulator, which by itself is a Bioconductor package in R\cite{frazee_polyester_2015}. This application was created under the namespace \texttt{Bio::SeqAlignment::Applications::Sequencing::Simulators::RNASeq::Polyester}.
It consists of two executable command line scripts: 
\begin{enumerate}
\item a Perl script (\texttt{polyester\_polyA.pl}) that parses  command line arguments with \texttt{Getopt::Long}. These include ALL the arguments that one would provide to the top level simulation function of the \texttt{Polyester} Bioconductor package from within R AND the parameters that control the stochastic simulation of the polyA tails.
\item the Perl application then calls a companion R script(\textit{polyester.R}) through the command line providing the arguments required by the simulation function in \texttt{Polyester},  generates the simulated sequencing data and writes them out to the disk. 
\end{enumerate}
Using the command line to communicate comes with certain benefits: a) the interface of the two scripts is modeled after the arguments of the top level simulation function in \texttt{Polyester}, so that the additional documentation could be kept to a minimum b) the option parsing packages in R, \texttt{getopt},  and Perl,  \texttt{Getopt::Long} are so similar, that an AI ChatBot can be used to generate the code to parse the argument list in one language from the argument parsing sequence in the other (in fact, we first wrote the R command-line API and had Github Copilot generate the Perl version of the argument parsing code).  The application itself is very simple, with a very simple pipeline structure which should, at some point, be re-written using only CPAN pipeline modules).

\begin{lstlisting}[language=Perl,basicstyle=\footnotesize,frame=single,caption={Pipeline logic of the \texttt{polyester\_polyA.pl} application.},captionpos=b]

## Pipeline Segment 1 : read sequences from the fasta file
my $bioseq_objects = read_fastx_sequences($fastafile);

## Pipeline Segment 2 : set up various defauls if not provided
...

## Pipeline Segment 3 : adds polyA tails to the sequences & record what was done
my $modifications_HoH =
  add_polyA( $bioseq_objects, $taildist, $seed, @distparams );

## Pipeline Segment 4 : write the modified sequences to a new fasta file
...

## Pipeline Segment 5 : simulate reads with R
Polyester_run(...);

## Pipeline Segment 6 : store the modifications into a file for future use
my $mod_fname = store_modifications(
    mods        => $modifications_HoH,
    bioseq_file => $fastafile,
    format      => $modformat
);

## Pipeline Segment 7 : split files into smaller files
...

## Pipeline Segment 8 : cleanup
unlink $readsfile unless $has_readfile;
unlink $fastafile;
\end{lstlisting}

To use the script one must provide a FASTA file of reference sequences (e.g. \textit{myseq.fasta}), the distribution (argument \textit{taildist}, one of the distributions in the GNU Scientific Library, GSL) that provides an appropriate model for the length of the polyA tail, the distribution parameters (including the left and the right truncation limits as the last elements of the \textit{distparams} list), and the data persistence format used to record the modifications (currently one of YAML, JSON or MessagePack). The remaining arguments are those that the \texttt{Polyester} expects; sensible default values are defined for most parameters and may be omitted from the command-line interface. An example invocation will look something like:

\begin{lstlisting}[language=bash,basicstyle=\footnotesize]
  polyester_polyA.pl --fastafile myseq.fasta --taildist gamma \
  --distparams 125.0 1.0 0.0 250.0 --fraglen 100 --fragsd 10  \
  --numreps 1 --strandspec TRUE --readlen 75 --paired F       \
  --maxseqs 1000 --modformat YAML --outdir /path/to/output
\end{lstlisting}

During the construction of the enhanced simulator we implemented several modules under the namespace\\ \texttt{Bio::SeqAlignment::Components}.  \Fref{tab:ModulesForPipeline}, shows a some of these modules that either support (e.g. read sequences using the BioX::Seq module) or are supported by the simulator. The complete list of modules may be found in the CPAN module \texttt{Bio::SeqAlignment::Examples::TailingPolyester}.

\begin{table}[!h]\footnotesize\centering
\caption[Table caption text]{Sample of modules supporting or supported by the Enhanced Polyester}
	\begin{tabularx}{1.0\textwidth}{X|p{6.55cm}|p{3.8cm}}
\toprule
	Package Name & Functionality & Implementation Details\\
\midrule
	\colorbox{black}{x}\texttt{::TrimTail} & Provides a generic interface to tail trimming operations. Pluggable module supporting IO from disk (traditional Unix filter) or from memory (mixed modes are also supported). & Inherits from the CPAN module \texttt{Moo}. \\
	\colorbox{black}{x}\texttt{::TrimTail::Regex} & Plugin for the TrimTail module that uses regular expression pattern matching to tril tails from sequences & Inherits from the CPAN module \texttt{Moo::Role}\\
\colorbox{black}{x}\texttt{::TrimTail::ChangePointCInline} & Plugin for \texttt{TrimTail} that applies statistical methods for changepoint detection to identify and trim polyA tails. & Inherits from the CPAN module \texttt{Moo::Role} and uses \texttt{Inline::C} to provide a naive C implementation for changepoint detection.\\
\colorbox{black}{x}\texttt{::Conversions::BioXFASTX} & Handles 
the conversion of in-memory lists of \texttt{BioX::Seq} objects to FASTX (where X is either A or Q
indicating a FASTA or a FASTQ) file in the disk. & Inherits from the CPAN module\texttt{Moo::Role}.\\
\bottomrule
\end{tabularx}
\vspace{12pt}
\raggedright\footnotesize{\colorbox{black}{x} stands for \texttt{Bio::SeqAlignment::Components}}
\label{tab:ModulesForPipeline}
\end{table}

\subsection{A Tiny Role for Randomness}
The enhanced simulator requires the user to specify the randomness they want to appear in the sequence tails, but how does the module itself generate random numbers under exact specifications? While one can use custom methods to simulate random numbers from (truncated) distributions, there is one truly generic method that can be applied to all distributions, i.e. the inverse CDF\footnote{As one may recall from introductory statistics, the CDF of a real-valued random variable X, evaluated at x, is the probability function that $X$ will take a value less than or equal to $x$} method. Given a random variable $X$ with CDF $F_X(x)$, the method is based on the principle that if $U$ is a uniform random variable on $(0,1)$, then $X = F^{-1}_X(U)$ has the same distribution as $X$.This method requires that the CDF $F_X(x)$ is invertible, which is the case for most continuous distributions. For discrete distributions, the method can be adapted by using the inverse of the discrete CDF. Here is what is the inverse CDF algorithm looks like: 

\begin{enumerate}
\item Generate a uniform random number $u$ in the interval $(0,1)$ for the non-truncated case. To sample from the distribution of a random variable $X$ that is truncated to the interval $[L,R]$ one draws the uniform random value in the interval $(F_X(L),CDF(F_XR))$ .
\item Compute $x = F^{-1}_X(u)$, where $F^{-1}_X(u)$ is the inverse of the CDF function\footnote{The inverse CDF function is also known as the quantile or the percentile function of the distribution} of $X$.
\item The number $x$ is a random number distributed according to the distribution of $X$.
\end{enumerate}

The necessary components for the algorithm are thus: a) methods to compute $F_X(x)$ and $F^{-1}_X(x)$ for an specific distribution $X$ and b) facilities to draw uniform random numbers. One reasonably clean way to pack this logic into a software component is to 
\begin{enumerate}
\item import all the relevant definitions from a module that provides methods for the CDF and inverse CDF methods for a \textit{very large} collection of random distributions into a new module
\item compose a role for the inverse CDF method into the module of the previous step
\item provide the module for the simulation with a plugin for the random number simulator
\end{enumerate}

With this design, one examine the performance gains of different codings of the inverse CDF method, i.e. by composing modules for stochastic simulation that compose alternatives roles AND different random number simulators. The latter is particularly important as there alternatives for the underlying engine that draws the random uniform numbers 
that differ in speed of generation, statistical correctness and (less relevant for our application) cryptographic security. In Perl, two modules \texttt{PDL::GSL::CDF} and \texttt{Math::GSL::CDF} provide interfaces to the CDF functions of the (very large) collection of random distributions in the GSL library. The main difference between the two packages is that the former provides access to these distributions via a \texttt{PDL} interface, while the latter is a pure Perl module. The \texttt{PDL} also provides alternatives to the Perl \texttt{rand} function for the uniform random number generator (RNG); one can also use the RNG from the GSL itself, or PDL's builtin replacement \texttt{random} that is based on the highly performant, but not cryptographically secure, Xoshiro generators \cite{blackman_scrambled_2021}, a family of RNGs with implementations in numerous programming languages \url{https://prng.di.unimi.it/} . \\
This is the role that provides the interface to the native Perl RNG:

\begin{lstlisting}[language=Perl,basicstyle=\footnotesize,frame=single,caption={A simple OO API for random number generation.},captionpos=b]
use strict;
use warnings;

package Bio::SeqAlignment::Examples::TailingPolyester::PERLRNG;
use Role::Tiny;

sub init { undef; }

sub random {
    my ( $self, $random_dim ) = @_;
    my $num_rand_values = 1;
    $num_rand_values *= $_ for @$random_dim;
    my @retvals = map { rand() } 1 .. $num_rand_values;
    \@retvals;
}

sub seed {
    my ( $self, $seed ) = @_;
    if ($seed) {
        $self->{seed} = $seed;
        srand($seed);
    }
    $self->{seed};
}
1;
\end{lstlisting}
The user can ask for random numbers over a number of dimensions and the random function will return them as a flat array reference. The interfaces to the PDL and the GSL\footnote{The GSL generator also requires an explicit initialization step, since that library supports multiple concurrent seeded generators by design} generators are defined analogously, i.e. they include a \texttt{random} and \texttt{seed} methods.
\\
The role for simulating, possibly truncated, distributions using vectorized PDL instructions is shown at \Fref{lst:RNGSimulator}. The role requires that any modules consuming it provide a few methods, i.e.  \texttt{random} for uniform random number generation,  \texttt{cdf}, \texttt{inv\_cdf} for the calculation of the CDF and the inverse CDF functions,and finally  \texttt{has\_distr} to check whether the distribution to be simulated from is provided  by the consuming module. The method \texttt{simulate\_truc} receives its methods through a \textit{hash}, and missing arguments are set to sensible default values. This allows us for example to return a single random number if the user does not specify how many they need, while also simulating left, right, and non-truncated versions if the relevant arguments are not provided by the caller. 

\begin{lstlisting}[language=Perl,basicstyle=\footnotesize,frame=none,caption={An API for the Inverse CDF method.},label={lst:RNGSimulator},captionpos=b]
use strict;
use warnings;

package Bio::SeqAlignment::Examples::TailingPolyester::SimulateTruncatedRNGPDL;

use Carp;
use PDL;
use Role::Tiny;
requires qw(random has_distr cdf inv_cdf);

sub simulate_trunc {
    my $self       = shift;
    my %args       = @_;
    my $random_dim = $args{random_dim} || [1];     ## default to 1 random number
    my $distr      = $args{distr}      || 'flat';
    my $left_trunc_lmt  = $args{left_trunc_lmt}  || 'missing';
    my $right_trunc_lmt = $args{right_trunc_lmt} || 'missing';
    my $params          = $args{params}          || 'missing';
    my $cdf_val_at_left_trunc_lmt;
    my $cdf_val_at_right_trunc_lmt;

    ## set up sanity checks
    croak "The distribution $distr is not available"
      unless $self->has_distr($distr);             ## distr must exist
    if ( $params eq 'missing' && $distr ne 'flat' ) {
        croak "Must provide parameters for the distribution $distr";
    }    ##and parameters cannot be missing unless it is a flat distribution

    ## set up CDF computations at the truncation limits
    if ( $left_trunc_lmt eq 'missing' ) {
        $cdf_val_at_left_trunc_lmt = pdl(0);
    }
    else {
        $left_trunc_lmt = pdl($left_trunc_lmt);
        $cdf_val_at_left_trunc_lmt =
          $self->cdf( $distr, $left_trunc_lmt, $params );

    }
    if ( $right_trunc_lmt eq 'missing' ) {
        $cdf_val_at_right_trunc_lmt = pdl(1);
    }
    else {
        $right_trunc_lmt = pdl($right_trunc_lmt);
        $cdf_val_at_right_trunc_lmt =
          $self->cdf( $distr, $right_trunc_lmt, $params );
    }
    my $domain_lengh_trunc_distr =
      $cdf_val_at_right_trunc_lmt - $cdf_val_at_left_trunc_lmt;

    ## now simulate the truncated distribution
    ...
}

1;
\end{lstlisting}
The API of the \texttt{cdf}, \texttt{inv\_cdf} is very simple: these methods receive the name of the distribution whose (inverse)CDF value is requested, followed by the value of the random value we are asking the calculation at and finally the parameters of the distribution as an array reference. 
As shown in \Fref{lst:InvCDFWithPDL}, coding the actual inverse CDF calculation using \texttt{PDL} takes four lines of code and can happen efficiently and with a single memory allocation because of PDL's capabilities for vectorized, in-place mathematical operations.
\pagebreak 
\begin{lstlisting}[language=Perl,basicstyle=\footnotesize,frame=none,caption={The inverse CDF calculations in  PDL.},label={lst:InvCDFWithPDL},captionpos=b]
    ## now simulate the truncated distribution
    my $simulated_values = $self->random($random_dim);
    $simulated_values->inplace->mult($domain_lengh_trunc_distr);
    $simulated_values->inplace->plus($cdf_val_at_left_trunc_lmt);
    return $self->inv_cdf( $distr, $simulated_values, $params );
\end{lstlisting}

In base Perl, one can either have their cake (use \texttt{map} to effectively vectorize the calculation, but forego in-place operations while paying the price of multiple memory allocations for each successive \texttt{map}) or eat it (use a \texttt{for} loop to modify things in place). For the comparator timings discussed below, we decided to eat the cake and opted for the C-like code shown in \Fref{lst:InvCDFWithPDL2}. This is the largest change one has to make to \Fref{lst:RNGSimulator} to code the inverse CDF method in base Perl. 

\begin{lstlisting}[language=Perl,basicstyle=\footnotesize,frame=none,caption={The inverse CDF calculations in base Perl.},label={lst:InvCDFWithPDL2},captionpos=b]
    ## now simulate the truncated distribution - non vectorized code
    my $simulated_values = $self->random($random_dim);
    my $num_rand_values  = scalar $simulated_values->@*;
    for my $i ( 0 .. $num_rand_values - 1 ) {
        $simulated_values->[$i] =
          $simulated_values->[$i] * $domain_lengh_trunc_distr +
          $cdf_val_at_left_trunc_lmt;
        $simulated_values->[$i] =
          $self->inv_cdf( $distr, $simulated_values->[$i], $params );
    }
    return $simulated_values;
\end{lstlisting}

The consumer of the previously defined roles, requires a collection of statistical distributions (\textit{collector}), but it's API is not affected by the choice of a specific collector. To keep things simple we hardwired collector in the consumer code. The version that consumes \Fref{lst:RNGSimulator} is shown at \Fref{lst:ConsumerPDL}. 

\begin{lstlisting}[language=Perl,basicstyle=\footnotesize,frame=none,caption={Consumer module for truncated random number generation.},label={lst:ConsumerPDL},captionpos=b]
use strict;
use warnings;

package Bio::SeqAlignment::Examples::TailingPolyester::SimulatePDLGSL;

use Package::Stash;
use PDL::Lite;
use PDL::GSL::CDF;
use subs        qw (seed  has_distr random is_rng_an_object);
use Class::Tiny qw(seed  has_distr seed ), { random => sub { } };
use Role::Tiny::With;
with 'Bio::SeqAlignment::Examples::TailingPolyester::SimulateTruncatedRNGPDL';

sub has_distr {
    my ( $self, $distr ) = @_;
    return exists $self->{distributions}->{$distr};
}

sub cdf {
    my ( $self, $distr, $lmt, $params ) = @_;
    return $self->{distributions}->{$distr}{cdf}->( $lmt, $params->@* );
}

sub inv_cdf {
    my ( $self, $distr, $lmt, $params ) = @_;
    return $self->{distributions}->{$distr}{inv_cdf}->( $lmt, $params->@* );
}

sub BUILD {
    ...
}

1;
\end{lstlisting}

Internally the consumer uses a two level hash (reference) for book-keeping: the key of the first level is the name of the distribution, and that of the second level is the \textit{type} of the function stored, i.e. one of "cdf" or "inv\_cdf". The \texttt{BUILD} method of the consumer illustrates a few features that reduce needless typing and increase flexibility and choice for the programmer to sugarcoat the hard-wiring of the collector\footnote{The code in \Fref{lst:ConsumerPDLBUILD} can easily be modified see e.g.  the CPAN documentation for \texttt{Package::Stash} to avoid hardwiring the collector.} :
\begin{enumerate}
\item the ingestion of the collector (in this case the module \texttt{PDL::GSL::PDF}), and the parsing of the name of the distributions from its own symbol table to populate the aforementioned hash structure, using \texttt{Package::Stash}.
\item the impossibility of ingesting the distributions without an consistent nomenclature for the latter in collector modules
\item the application of the random number generator at \textit{runtime} using an RNG plugin with appropriate initialization and seeding
\end{enumerate}

\begin{lstlisting}[language=Perl,basicstyle=\footnotesize,frame=none,caption={BUILD method for the consumer module for truncated random number generation.},label={lst:ConsumerPDLBUILD},captionpos=b]
sub BUILD {
    my ( $self, $args ) = @_;
    die "seed must be a number" unless $self->{seed} =~ /^[0-9]+$/;
    my @symbols_in_GSL =
      Package::Stash->new('PDL::GSL::CDF')->list_all_symbols('CODE');
    for (@symbols_in_GSL) {
        if (/gsl_cdf_(\w+)_Pinv/) {
            $self->{distributions}->{$1} = {
                cdf     => $PDL::GSL::CDF::{"gsl_cdf_$1_P"},
                inv_cdf => $PDL::GSL::CDF::{$_}
            };
        }    ## store the CDF and inverse CDF functions
    }
## initialize and store the RNG here
    if ( $args->{RNG_init_parameters} ) {
        $self->{is_rng_an_object} = 1;
        $self->{rng_object} =
          $args->{rng_plugin}->init( $args->{RNG_init_parameters} );
    }
    die "RNG plugin does not provide the 'random' role"
      unless $args->{rng_plugin}->can('random');
    Role::Tiny->apply_roles_to_object( $self, $args->{rng_plugin} );
}
\end{lstlisting}

The code below shows how one can use the consumer API: first one initializes a random number generator (a pattern that should be familiar to users of the GSL library or Intel's oneAPI library for machine learning) and then simulates from a given distribution (in this case the lognormal one with mean parameter $log(125)$, scale parameter $1$, truncating it to the interval $[0,250]$.
\medskip 

\begin{minipage}{0.44\textwidth}
\begin{lstlisting}[language=Perl,basicstyle=\footnotesize,frame=single,captionpos=b]
my $distr             = 'lognormal';
my $params            = [ log(125), 1 ];
my $lower_trunc_limit = 0;
my $upper_trunc_limit = 250;
my $samples            = 1000000;
\end{lstlisting}
\end{minipage}\hfill
\begin{minipage}{0.50\textwidth}
\begin{lstlisting}[language=Perl,basicstyle=\footnotesize,frame=single,captionpos=b]
my $rng = SimulatePDLGSL->new(
    seed       => 3, 
    rng_plugin => 'PERLRNGPDL');
my $pdl = $rng->simulate_trunc(
    random_dim      => [$samples],
    distr           => $distr,
    params          => $params,
    left_trunc_lmt  => $lower_trunc_limit,
    right_trunc_lmt => $upper_trunc_limit);
\end{lstlisting}
\end{minipage}
\medskip

\begin{figure}[H]
\centering  
\includegraphics[width=\textwidth]{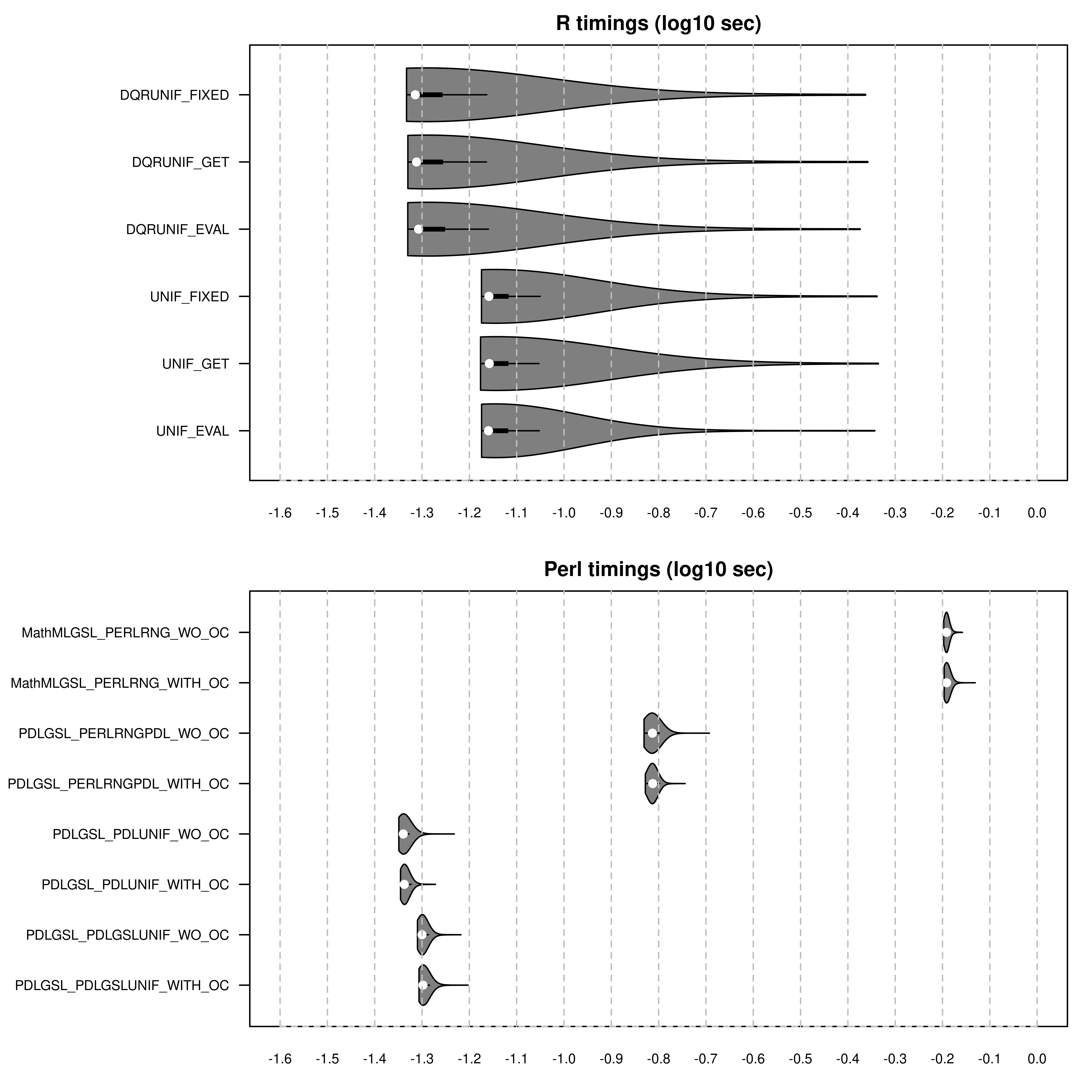}
\caption{Performance evaluation of Random Number Generation in R (upper panel) and Perl (lower panel). The figure shows violin plots of 1,000 replicates of a task that involves simulating one million random numbers from the log-normal distribution with mean parameter $log(125)$, scale parameter $1$, truncated to the interval $[0,250]$. The performance of the Xoshiro RNG (DQRNG in R, PDLUNIF in Perl) was evaluated vis-a-vis the builtin RNGs (UNIF in R, PERLRNG in Perl) and a the uniform RNG from GSL in Perl). Other scenarios evaluated included: a) the impact of object construction (WO\_OC used a single object for all 1,000 replications v.s. WITH\_OC that created a new object for each replicate), b) using non-vectorized base Perl for RNG (PERLRNG) instead of PDL (PDLUNIF) c) using non-vectorized, base Perl, for the inverse CDF method (MATHGSL) and d) alternatives in R to provide dependencies such as the RNG and the (inverse)CDF functions of the target distribution \Fref{apndx:RNGCodeInR}}\label{fig:LognormalRNGTiming}
\end{figure}

At this point one may question the complexity of the design with separate roles for the inverse CDF algorithm and the uniform RNG, and a consumer module that wraps over statistical distributions for consumption. Wouldn't it be much simpler to just code the "meat" code \texttt{simulate\_trunc} as a subroutine, provide two code references as arguments and get it over and done with? Or even use an \texttt{eval} string, or just hard wire the random number simulators? Indeed, any of the above methods would have been much simpler to write, but it would have made a performance evaluation of \texttt{PDL} v.s. base Perl, or comparisons among different RNGs more verbose to write and more complicated to understand. 
The script \texttt{testRN\_performance.pl} found under the \texttt{scripts} folder in the \texttt{Bio::SeqAlignment::Examples::TailingPolyester} CPAN module, evaluates 8 different combinations to simulate from the lognormal distribution using the API we presented in fewer than 200 lines (most of the "lines" are just there for formatting purposes and/or to set up recording of the execution times of the alternatives to truncted distribution simulation). 
To undertake a cross-language performance evaluations we turned to the language R, which has a very similar to Perl approach for managing entries in symbol tables, while also having access to an implementation of the Xoshiro family of RNGs available in the package \texttt{dqrng}\footnote{\url{https://github.com/daqana/dqrng}}. In \Fref{apndx:RNGCodeInR} we provide code for these three "easier" implementations, i.e. obtaining functions through symbolic reference search over the visible stashes\footnote{This is accomplished via the R function \texttt{get} which can do a search for a symbol entry in a specific \texttt{environment}, or all the environments that are within scope at the current point in the code.  The environment is R's version of a Perl's package.}, eval strings, or plain old fixing of the name of the dependency in the code. In the absence of checks for valid arguments the code is indeed shorter, but the part of the code that does the actual calculations (last four lines in each function at \Fref{apndx:RNGCodeInR}) is exactly the same length as the PDL code in \Fref{lst:InvCDFWithPDL}.
\\
In \Fref{fig:LognormalRNGTiming} we show violin plots, i.e. smoothed histograms of absolute timings for generating 1 million truncated lognormal random variables using six different implementations in R and 8 in Perl. To assess variability in performance, we generated one thousand replicates of this task for each implementation in either language. There are a few noteworthy observations to be made about this data:
\begin{enumerate}
\item Lack of vectorization is the major impediment to high performance Perl scripts for numerical or statistical calculations for big data. This is clearly seen when comparing the performance of the 1) MathMLGSL\_PERLRNG implementations that use base Perl for  both random number generation and for coding of the inverse CDF method by \Fref{lst:InvCDFWithPDL2} to that of 2) PDLGSL\_PERLRNG that generate random numbers through base Perl, but use the PDL code in \Fref{lst:InvCDFWithPDL} nd 3) the pure PDL implementations PDLGSL\_PDLUNIF.  
\item The speed of the vectorized \texttt{PDL} implementation that also used the \texttt{PDL} RNG were competitive with the R code that used the Xoshiro implementation.
\item The Xoshiro based implementations achieved the best performance in either language. In R the difference is largely due to the speed of the RNG relative to the one builtin the language, whereas in Perl the difference can be attributed to the non-vectorized nature of base Perl and the RNG. The former is the biggest source of performance drop as can be seen from the comparison of the lower four violin plots in the lower panel of \Fref{fig:LognormalRNGTiming}, which swapped the Xoshiro for the GSL default RNG, but kept vectorization for calculations. 
\item Object construction (contrast violin plots "WO\_OC" vs "WITH\_OC") had negligible effect on the relative performance of all Perl implementations. 
\item Despite appeals to avoid "eval" strings in R, the choice of method to access dependencies (e.g. the inverse CDF method, or the RNG) had minimal effect on performance. 
\end{enumerate}
\textbf{In summary, these observations should prompt a re-examination of the combination of Perl and PDL in data analytics.} The \texttt{PDL} packages offer a bewildering array of performant modules for numerical and statistical calculations that nicely complement Perl's metaclass programming facilities. 
\newpage
\newpage

\subsubsection{Trimming the polyA tail from RNA sequences }
After all the trouble we went to attach polyA tails to sequences, why should we even want to remove them? As noted in \Fref{sec:RNADNAsequencing}, such tails are features introduced via nature (messenger RNA) or protocol (e.g. our PALS-NS) to the "text" that comes out of the genome. The basic analysis task for any RNAseq experiment is the mapping of the sequence generated by the experiment back to the genome that does not have polyA tails tucked at the end of each sequence. Hence, while \textit{adding} polyA tails is a pre-requisite for a realistic sequence simulation, \textit{removing} the polyA tail is very important for accurate sequence mapping; as the latter is always done via an approximate text matching approach, having extra text in the query may, and in fact will, lead to inaccurate mappings. 
\\
The current state-of-the-art method for removing polyA tails is the \texttt{cutadapt} Python program\cite{martin_cutadapt_2011} that was introduced for 2nd generation, short RNA sequencing more than 10 years ago. That program has numerous other functionalities (e.g. removing adapters found as decorator features in sequences), but for our purposes the relevant one  is its ability to trim polyA tails. The most recent version of the polyA trimming algorithm\footnote{\url{https://cutadapt.readthedocs.io/en/stable/algorithms.html\#poly-a-algorithm}} in cutadapt is intrinsically a sequence alignment. This intrinsic alignment is based on an  scoring scheme that awards a +1 for an A letter and  -2 for a non-A letter at a given position. The algorithm considers all possible suffixes in the sequence of interest, and after filtering those that have more than 20\% non-A letters, returns the position of the suffix with the largest score as the beginning of the tail. Note that the alignment between the sequence of interest and the is never explicitly formed, thus the algorithm avoids the $O(N^2)$ scaling of a bona fide alignment method. 
The Python code for the polyA implemented by \texttt{cutadapt}  for an arbitrary string \textit{s} is shown in \Fref{lst:CutadaptInPython}. Due to the linear relationship between the number of errors and the best score, it is sufficient to test for the proportion of errors \textbf{after} the best index has been found. This leads to a modified \texttt{cutadapt} whose Perl code is shown in \Fref{lst:ModifiedCutadaptInPerl}.

\noindent\hspace{0.20\linewidth}\begin{minipage}{0.60\textwidth}
\begin{lstlisting}[language=Python,basicstyle=\footnotesize,frame=none,caption={Python code for \texttt{cutadapt}.},label={lst:CutadaptInPython},captionpos=b]
n = len(s)
best_index = n
best_score = score = errors = 0
for i, nuc in reversed(list(enumerate(range(n)))):
    if nuc == "A":
        score += 1
    else:
        score -= 2
        errors += 1
    if score > best_score and errors <= 0.2 * (n - i):
        best_index = i
        best_score = score
s = s[:best_index]
\end{lstlisting}
\end{minipage}

\noindent\hspace{0.20\linewidth}\begin{minipage}{0.60\textwidth}
\begin{lstlisting}[language=Perl,basicstyle=\footnotesize,frame=none,caption={Perl code for a modified \texttt{cutadapt}.},label={lst:ModifiedCutadaptInPerl},captionpos=b]
my $n          = length $s;
my $best_index = $n;
my $best_score = my $score = 0;
foreach my $i ( reverse( 0 .. $n - 1 ) ) {
    my $nuc = substr $s, $i, 1;
    $score += $nuc eq 'A' ? +1 : -2;
    if ( $score > $best_score ) {
        $best_index = $i;
        $best_score = $score;
    }
}
$best_index = $n - $best_index;
if ( $best_score < 0.4 * ( $best_index + 1 ) ) {
    $best_index = $n;
}
substr( $s, -$best_index ) = '';
\end{lstlisting}
\end{minipage}

\newpage
However, this is not the only possible Perl implementation for a trimming method. A few other possibilities do arise at this point:
\begin{enumerate}
\item use a \texttt{cutadaptMAP} rather than a loop to calculate the best scoring tail as in \Fref{lst:CutadaptMap}
\item code the loop that searches for the position with the highest score in a vectorized PDL operation (\texttt{cutPDL}) in \Fref{lst:CutadaptInPDL}
\item use the package \texttt{Inline::C} to wrap around a pure C implementation of the (modified) \texttt{cutadapt} method (\texttt{cutC})
\item use a regular expression to match for a suffix that is guaranteed to have a percentage of A letter that
is over 25\% as in \Fref{lst:CutadaptAsRegex}
\item use \texttt{Inline::C} to wrap around a changepoint detection method. In the latter we compute a statistical test that the tail begins at a given position against the \textit{null} hypothesis that no tail exists in the sequence of interest. To do so, we form simple composition models for the letters in the header and the tail of the sequence that are based on flipping biased "coins", with heads being a non-A letter and tails being an A letter. We then compute the maximum of the likelihood of having a tail sequence over all possible positions. If the maximum likelihood of having a polyA tail is greater than the likelihood of a "tail-less" sequence by a given, user defined threshold, e.g. a 1:1000 odds, then we use report the position that the maximum likelihood was observed as the beginning of the polyA tail. 
\end{enumerate}
\noindent\hspace{0.15\linewidth}\begin{minipage}{0.70\textwidth}
\begin{lstlisting}[language=Python,basicstyle=\footnotesize,frame=none,caption={A \texttt{map} for \texttt{cutadapt}.},label={lst:CutadaptMap},captionpos=b]
my $s = "ACTGCAAAAAAAA";  ## test string  
my $n          = length $s;
my $sum        = 0;
my $best_score = my $score = 0;
my @s =
    map { $sum += ( $_ eq 'A' ? +1 : -2 ) }
    split //, reverse $s;
my $best_index = reduce { $s[$a] > $s[$b] ? $a : $b } 0 .. $#s;
$best_index =
    ( $s[$best_index] < 0.4 * ( $best_index + 1 ) )
    ? $n
    : $n - $best_index - 1;
$s = substr $s, 0, $best_index;
    
\end{lstlisting}
\end{minipage}

\noindent\hspace{0.08\linewidth}\begin{minipage}{0.84\textwidth}
\begin{lstlisting}[language=Perl,basicstyle=\footnotesize,frame=none,caption={Perl code for a modified \texttt{cutadapt} that uses \texttt{PDL}.},label={lst:CutadaptInPDL},captionpos=b]
use PDL;
use PDL::Primitive qw (where_both);
use PDL::Ufunc     qw (cumusumover);
use constant ascii_A => ord('A');

my $s = "ACTGCAAAAAAAA";  ## test string   
my $pdl_s = PDL->new( [ unpack( "C*", reverse $s ) ] );
my ( $pdl_index, $pdl_c_index ) = where_both( $pdl_s, $pdl_s == ascii_A );
$pdl_index   .= 1;
$pdl_c_index .= -2;
my $score      = cumusumover($pdl_s);
my $max_index  = $score->maximum_ind;
my $best_index = $n - $max_index - 1;
if ( $score->at( $score->maximum_ind ) < 0.4 * ( $max_index + 1 ) ) {
    $best_index = $n;
}
$s = substr $s, 0, $best_index;
\end{lstlisting}
\end{minipage}

\noindent\hspace{0.15\linewidth}\begin{minipage}{0.70\textwidth}
\begin{lstlisting}[language=Python,basicstyle=\footnotesize,frame=none,caption={A regex for \texttt{cutadapt}.},label={lst:CutadaptAsRegex},captionpos=b]
my $polyA_min_25_pct_A = qr/
                ( ## match a poly A tail which is 
                  ## delimited at its 5' by *at least* one A
                  A{1,}
                      ## followed by the tail proper which has a    
                  (?:     ## minimum composition of 25% A, i.e.
                      ## we are looking for snippets with 
                      (?: 
                              ## up to 3 CTGs followed by at 
                              ## least one A
                              [CTG]{0,3}A{1,}    
                      )
                      |     ## OR
                      (?: 
                              ## at least one A followed by 
                              ## up to 3 CTGs
                              A{1,}[CTG]{0,3}
                      )
                  )+    ## extend as much as possible
              )\z/xp;

my $s = "ACTGCAAAAAAAA";  ## test string              
my $n          = length $s;
$s =~ m/$$polyA_min_25_pct_A/;
my $best_index = length $1;
substr( $s, -$best_index ) = '';
    
\end{lstlisting}
\end{minipage}

Benchmarking of these potential alternatives (\fref{tab:PolyATrimmingPerformance}) shows that there is always a Perl solution that delivers a performance similar to that of the gold standard \texttt{cutadapt}; in fact there are multi-language solutions that are orders of magnitude faster than that Python application.  This brings us to the next question: how can we serve components for data intensive work flows to the user in a "black-box" manner? And more importantly, what are some potential advantages that Perl brings to this table?

\setlength{\footnotesep}{0.5cm}
\begin{table}[!h]\centering
\caption[Table caption text]{Time to trim a polyA tail that is 20\% of the length of a target sequence}
	\begin{tabularx}{1.0\textwidth}{YYYYYY@{}}
\toprule
Algorithm & Language & \multicolumn{4}{c}{Target Sequence Length}\\
\hline\\
	\hspace{1pt} & \hspace{1pt} & 100     &     1000        &   2000         & 10000\\
	\cmidrule(r){3-6} \\
cutadapt       & Python  & $10.3\pm3.0$   & $103.0\pm10.0$  & $213.0\pm12.9$ & $1330.0\pm64.2$\\
cutadapt(mod)  & Python  & $8.1\pm1.8$    & $79.6\pm5.4$    & $163\pm9.5$    & $1130.0\pm86.8$\\   
cutadapt(mod)  & Perl    & $16.0\pm2.5$   & $150.0\pm11.1$  & $310.0\pm21.0$ & $1500.0\pm88.0$\\
catadaptMAP    & Perl    & $19.0\pm4.4$   & $180.0\pm17$    & $380.0\pm25$   & $1900.0\pm18.0$\\
cutPDL         & Perl    & $100.0\pm18.0$ & $190.0\pm13.0$  & $270.0\pm16.0$ & $1000.0\pm45.0$\\
regex          & Perl    & $26.0\pm10.0$  & $310.0\pm26.0$  & $620.0\pm42.0$ & $3200.0\pm140.0$\\
cutC           & Perl/C  & $0.6\pm1.0$    & $3.1\pm1.1$     & $6.0\pm1.4$    & $28.0\pm4.3$\\
changepoint    & Perl/C  & $3.6\pm2.0$    & $33.0\pm5.2$    & $65.0\pm8.2$   & $320.0\pm24.0$\\
\bottomrule

 \end{tabularx}
\vspace{12pt}
\raggedright\footnotesize{Performance is reported as mean $\pm$ standard deviation (in microsceconds) over 2000 repetitions}
\label{tab:PolyATrimmingPerformance}
\end{table}

\newpage

\subsection{Enhancing Edlib with Meta-Object Programming (MOP) techniques}
\subsubsection{Data flows for sequence mapping applications}
\begin{figure}[h!]
		\centering
		\subfloat[LinearLinear Data Flow]{
			\begin{tikzpicture}[node distance=1cm, auto]
				\tikzset{
					box/.style={
						rectangle, 
						draw, 
						minimum width=2.5cm, 
						minimum height=0.8cm, 
						text centered
					},
					longbox/.style={
						rectangle,
						draw,
						minimum width=2.5cm,
						minimum height=5.0cm, 
						text width=2cm,
						align=center
					}
				}
				\node[longbox,label=\texttt{\textbf{Aligner}}] (alignerLL) {};
				\node[longbox, right=of alignerLL,label = \texttt{\textbf{Mapper}}] (mapperLL) {};
				
				\node[box, below left= 0.3cm and 1cm of alignerLL.north west] (query1LL) {Query Sequence 1};
				\node[box, below=of query1LL] (query2LL) {Query Sequence 2};
				\node[box, below=of query2LL] (querynLL) {Query Sequence n};
				
				\draw[->] ([yshift=0ex]query1LL.east) -- ([yshift=11.9ex]alignerLL.west);
				\draw[->] ([yshift=0ex]query2LL.east) -- ([yshift=0ex]alignerLL.west);
				\draw[->] ([yshift=0ex]querynLL.east) -- ([yshift=-11.9ex]alignerLL.west);
				\draw[->] ([yshift=11.9ex]alignerLL.east) -- ([yshift=11.9ex]mapperLL.west);
				\draw[->] ([yshift=0ex]alignerLL) -- ([yshift=0ex]mapperLL);
				\draw[->] ([yshift=-11.9ex]alignerLL.east) -- ([yshift=-11.9ex]mapperLL.west);	
				
				\draw[dashed] ([yshift=11.9ex]alignerLL.east) -- ([yshift=11.9ex]alignerLL.west);		
				\draw[dashed] ([yshift=0ex]alignerLL.east) -- ([yshift=0ex]alignerLL.west);	
				\draw[dashed] ([yshift=-11.9ex]alignerLL.east) -- ([yshift=-11.9ex]alignerLL.west);
				
				\draw[dashed] ([yshift=11.9ex]mapperLL.east) -- ([yshift=11.9ex]mapperLL.west);		
				\draw[dashed] ([yshift=0ex]mapperLL.east) -- ([yshift=0ex]mapperLL.west);	
				\draw[dashed] ([yshift=-11.9ex]mapperLL.east) -- ([yshift=-11.9ex]mapperLL.west);				
				
				\node[box,  below right=0.3cm and 2.2cm of mapperLL.north] (output1LL) {Best Match Sequence 1};
				\node[box, below=of output1LL] (output2LL) {Best Match Sequence 2};
				\node[box, below=of output2LL] (outputnLL) {Best Match Sequence n};
				
				\draw[->] ([yshift=11.9ex]mapperLL.east) -- (output1LL);
				\draw[->] ([yshift=0ex]mapperLL.east) -- (output2LL);
				\draw[->] ([yshift=-11.9ex]mapperLL.east) -- (outputnLL);
			\end{tikzpicture}
			\label{fig:linearlinear}
		}
		
		\vspace{1cm} 
		
		\subfloat[Linear Data Flow]{
			\begin{tikzpicture}[node distance=1cm, auto]
				\tikzset{
					box/.style={
						rectangle, 
						draw, 
						minimum width=2.5cm, 
						minimum height=0.8cm, 
						text centered
					},
					longbox/.style={
						rectangle,
						draw,
						minimum width=3cm,
						minimum height=5cm, 
						text width=2cm,
						align=center
					}
				}
				
				\node[longbox, label=\texttt{\textbf{Aligner/Mapper}}] (alignerMapperL) {};
				\node[box, below left= 0.25cm and 1cm of alignerMapperL.north west] (query1LL) {Query Sequence 1};
				\node[box, below=of query1LL] (query2LL) {Query Sequence 2};
				\node[box, below=of query2LL] (querynLL) {Query Sequence n};
				
				\draw[->] ([yshift=0ex]query1LL.east) -- ([yshift=12.2ex]alignerMapperL.west);
				\draw[->] ([yshift=0ex]query2LL.east) -- ([yshift=0.2ex]alignerMapperL.west);
				\draw[->] ([yshift=0ex]querynLL.east) -- ([yshift=-11.8ex]alignerMapperL.west);
				
				\node[box,below right=0.3cm and 2.6cm of alignerMapperL.north] (output1L) {Best Match Sequence 1};
				\node[box, below=of output1L] (output2L) {Best Match Sequence 2};
				\node[box, below=of output2L] (outputnL) {Best Match Sequence n};
				
				\draw[->] ([yshift=12.2ex]alignerMapperL.east) -- (output1L);
				\draw[->] ([yshift=0.2ex]alignerMapperL.east) -- (output2L);
				\draw[->] ([yshift=-11.8ex]alignerMapperL.east) -- (outputnL);
				
				\draw[dashed] ([yshift=12.2ex]alignerMapperL.west) -- ([yshift=12.2ex]alignerMapperL.east);
				\draw[dashed] ([yshift=0.2ex]alignerMapperL.west) -- ([yshift=0.2ex]alignerMapperL.east);
				\draw[dashed] ([yshift=-11.8ex]alignerMapperL.west) -- ([yshift=-11.8ex]alignerMapperL.east);				
			\end{tikzpicture}
			\label{fig:linear}
		}
		
		\caption{Data flows for sequence mapping}
		\label{fig:data_flows}
	\end{figure}
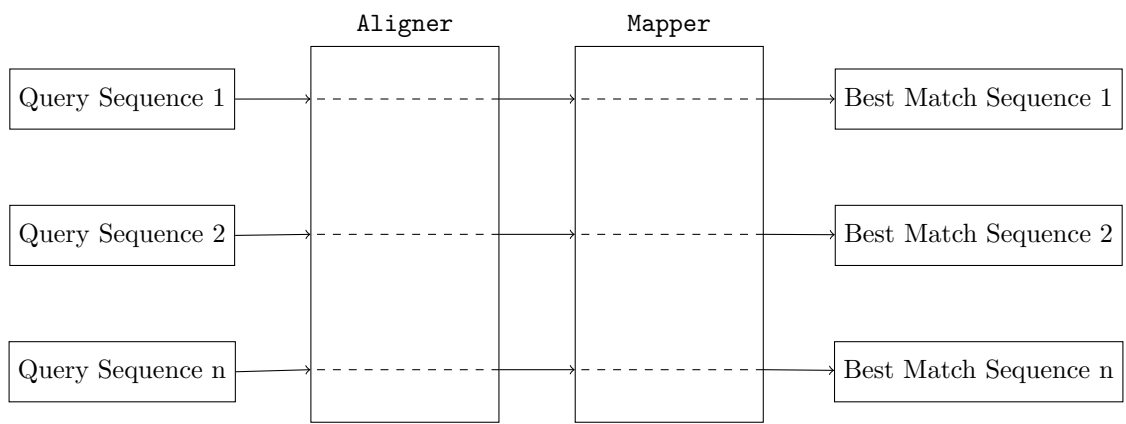
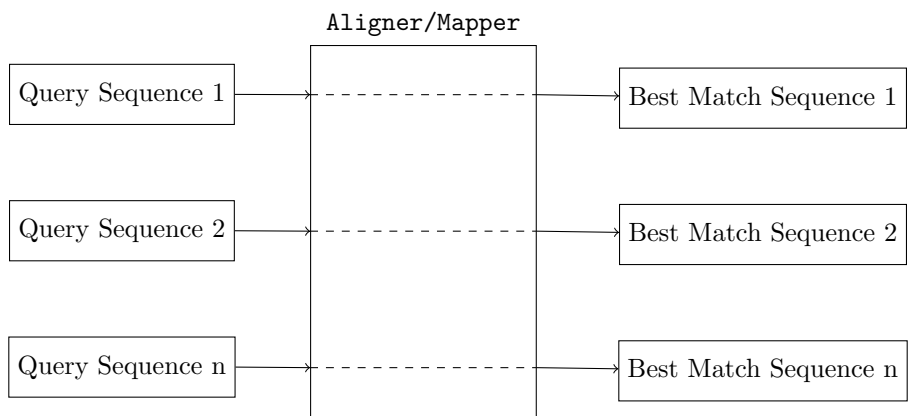

Identifying ("mapping") a query sequence with a reference one, is a task that follows a limited number of potential data flows, with two patterns (shown in \Fref{fig:data_flows}) accounting for most cases. In the first data flow, query sequences move independently through the component "\textbf{aligner}" that conducts the similarity search against the reference database via an exact alignment method (with or without heuristic speedups) or an approximate pseudo-alignment approach\cite{bray_near-optimal_2016}. The aligner yields a potentially filtered vector of similarity metrics (e.g. edit distance, similarity scores or p-values) for each sequence that the data flow controller directs to the second component, the "\textbf{mapper}". The latter carries out a \textbf{reduction operation} of the similarity vector into a single value, which is the considered the best match of the query to the universe of the reference sequences. In the second, data flow, the mapper and the aligner functionalities are carried out by the same component. The aligner and/or mapper components operate on the sequences in a linear, parallel fashion, i.e. there is no cross talk between alignment and mapping for any query sequences.  The two data flows are thus designated as \textbf{LinearLinear} and \textbf{Linear} respectively and either provides unique opportunities for parallelization, irrespective of the similarity metric or the reduction operation used in a given application.

\newpage
\subsubsection{Writing generic mappers with \texttt{Moose} and roles}
When developing a mapper for a given bioinformatic application, there are several potential alternatives to consider: in the \textbf{LinearLinear} case, one will typically have a choice among various aligners, similarity metrics and reduction methods, while in the \textbf{Linear} case this cornucopia of choices will be present, but hidden in the single aligner/mapper component. For a given application one may also want to compare implementation that follow either a \textbf{LinearLinear} or \textbf{Linear} further expanding the set of potential alternatives that one may wish to explore. One could create (and manage) a potentially very large collection of modules, one for each data flow and combination of components for each data flow, but one would most certainly would like to avoid this path, even if these modules are only created for internal use. A cultural feature should also be considered here: in the era of the "Jupyterization" of scientific and data intensive programming, attracting new users to a language requires that most experimental development takes place in a single file.  
One potential solution to managing the complexity of the number of modules and offering a single-file experience is by leveraging \texttt{Moose} MOP capabilities to create a generic sequence mapping module, while using roles to assign a particular data flow to a given generic mapper. Such a decomposition is handled by the \texttt{Bio::SeqAlignment::Components::SeqMapping} module that we created for this particular line of work (\Fref{tab:ModulesForSeqMapping}). 
The module also provides data flow roles for composition by fully fledged sequence mapping modules, that do not use the generic mapper interface. These data flow modules are nearly identical to their generic counterparts, with the only difference being the calling sequence of functions provided by the mapper that composes them into a role. The non-generic modules hopefully come out as a result of using the generic mapper module to develop the code for the alignment/mapping steps before releasing a  mature application to the world. This provides a somewhat elegant solution to the problem of exploration of sequence mapping component alternatives without having to create separate modules for each particular combination that failed some form of internal control. 

\begin{table}[H]\footnotesize\centering
\caption[Table caption text]{Modules supporting sequence mapping}
	\begin{tabularx}{1.0\textwidth}{X|p{6.0cm}|p{3.8cm}}
\toprule
	Package Name & Functionality & Implementation Details\\
\midrule
	\colorbox{black}{x}\texttt{::Mapper::Generic} & Provides a generic sequence mapping interface. & Inherits from the CPAN module \texttt{Moose}. \\
\colorbox{black}{x}\texttt{::Dataflow::LinearLinearGeneric} & A role for \texttt{Mapper::Generic} that supports the \textbf{LinearLinear} dataflow. & Inherits from  the CPAN module \texttt{Moose::Role}.\\
\colorbox{black}{x}\texttt{::Dataflow::LinearGeneric} & A role for \texttt{Mapper::Generic} that supports the \textbf{Linear} dataflow. & Inherits from the CPAN module \texttt{Moose::Role}.\\
\colorbox{black}{x}\texttt{::Dataflow::LinearLinear} & A role that supports the \textbf{LinearLinear} dataflow. It is intended to be composed into fully fledged sequence mapping modules, i.e. non-generic ones.  & Inherits from  the CPAN module \texttt{Moose::Role}.\\
\colorbox{black}{x}\texttt{::Dataflow::Linear} & A role that supports the \textbf{Linear} dataflow. It is intended to be composed into fully fledged sequence mapping modules, i.e. non-generic ones. & Inherits from the CPAN module \texttt{Moose::Role}.\\
\bottomrule
\end{tabularx}
\vspace{12pt}
\raggedright\footnotesize{\colorbox{black}{x} stands for \texttt{Bio::SeqAlignment::Components}}
\label{tab:ModulesForSeqMapping}
\end{table}

\noindent The \textbf{LinearLinearGeneric} role module requires that the module that consumes their role provide the following methods:
\begin{description}
\item[seq\_align] : this is the method that carries out the (pseudo)sequence alignment task for each work element.
\item[extract\_sim\_metric] : a method that extracts the similarity metric for each work element. 
\item[reduce\_sim\_metric] : a method that reduces the similarity metric to a single value for each work element.
\item[cleanup]: a method that handles cleanup operations, either in memory or in the disk depending on the space one is working in.
\end{description}

\noindent The \textbf{LinearGeneric} module require that the module consuming them provides the following methods:
\begin{description}
\item[seq\_align] : this is the method that carries out the method that carries out the (pseudo)alignment mapping and reduction tasks for each work element.
\item[cleanup]: a method that handles cleanup operations, either in memory or in the disk depending on the space one is working in.
\end{description}

\noindent  These modules assume that the work to be handled by the mapping module that consumes them will be provided as a reference to an array of distinct \textit{work elements}. The latter can be individual sequences, e.g. \texttt{Bio:Seq}, or FASTA files with many such sequences, or anything that the user feels appropriate for the space (in-memory or in-disk) they want their mapping application to work in. The consuming modules themselves don't care as long as the work elements are grouped together in an array, and the relevant reference is provided through the role API. The role modules provide a single method \texttt{sim\_seq\_search} , which is the role that is consumed by the (generic) mapper module.
\\
The generic mapper is a Moose module with pure data (hash reference) attributes, as well as attributes that are code references. The data attributes provide convenient storage and book keeping (e.g. which attributes have been set to non-default values) for the code reference arguments.  The names of the attributes of the generic mapper module are summarised in \Fref{tab:GenericMapperModule}. An \textit{around} modifier is applied to these arguments by the module as in  \Fref{lst:AroundModifier} with the explicit purpose of providing the object reference as the first argument to each method. This not only allows each method direct access to the other attributes of the object, but allows one to write the code corresponding to each attribute as if one were writing an instance method, rather than a code reference attribute for an object (or a hook in an non-OO setting). In the first case \texttt{\$self} would have been the first parameter passed, but in the second case, the function would have no \texttt{\$self} awareness. 
\begin{table}[!h]\centering
\caption[Table caption text]{Attributes of the \texttt{Bio::SeqAlignment::Components::Mapper::Generic} generic mapper module. }
	\begin{tabularx}{1.0\textwidth}{Y|Y|Y}
\toprule
Searching/Mapping Methods       & Reference Database Methods  & Parameters\\
\midrule\\
\texttt{init\_sim\_search}      & \texttt{create\_refDB}      & \texttt{refDB\_access\_params}\\
\texttt{seq\_align}             & \texttt{use\_refDB}         & \texttt{sim\_search\_params}\\
\texttt{extract\_sim\_metric}   &                             & \texttt{extract\_sim\_metric\_params}\\
\texttt{reduce\_sim\_metric}    &                             & \texttt{seqmap\_params}\\
\texttt{cleanup}                &                             & \texttt{\_has\_nondefault\_value}\\                       
\texttt{ }                      &                             & \texttt{\_code\_for}\\
\bottomrule
 \end{tabularx}
\label{tab:GenericMapperModule}
\end{table}

\noindent\hspace{0.00\linewidth}\begin{minipage}{1.0\textwidth}
\begin{lstlisting}[language=Perl,basicstyle=\footnotesize,frame=none,caption={Around modification of the methods of the generic mapper module.},label={lst:AroundModifier},captionpos=b]
around qw(
  create_refDB  use_refDB cleanup reduce_sim_metric 
  seq_align     init_sim_search   extract_sim_metric) => sub {
    my $orig = shift;
    my $self = shift;
    return sub {
        return $self->$orig->( $self, @_ );
    };
\end{lstlisting}
\end{minipage}

\noindent The code for the \textbf{LinearLinearGeneric}  in \Fref{lst:LinearLinearGenericRoleCode} implements process level parallelism through the \texttt{MCE} (Many Cores Engine) module. The final argument to the role method is a hash that provides key-value arguments that control this parallelism: the number of workers (if only one, the MCE will not be used) and the chunk size of the workload that each worker will be processing. The code for \textbf{LinearGeneric} is derived entirely analogously, with the only difference being the assumption that the \texttt{seq\_align} function will not only carry out the alignment, but also extract the vector of similarity metrics for each sequence and carry out any necessary reductions. \textbf{If any of the functions \texttt{seq\_align}, \texttt{extract\_sim\_metric} and \texttt{reduce\_sim\_metric} provide thread level parallelism, then the module consuming these roles will be endowed with both process and thread parallelism.} Both roles should deliver to the module that consumes them the final mapping between a query sequence and the reference sequence that is most similar to it, along with the value of the metric that was used to make this determination.

\newpage
\begin{lstlisting}[language=Perl,basicstyle=\footnotesize,frame=none,caption={Perl code for the \texttt{sim\_seq\_search} of the \textbf{LinearLinearGeneric} role.},label={lst:LinearLinearGenericRoleCode},captionpos=b]
package Bio::SeqAlignment::Components::SeqMapping::Dataflow::LinearLinearGeneric;
use strict;
use warnings;

use Moose::Role;
use Carp;
use MCE;
use MCE::Candy;
use namespace::autoclean;

requires 'seq_align';          ## the method that does (pseudo)alignment
requires 'extract_sim_metric'; ## extract the similarity metric from the search
requires 'reduce_sim_metric';  ## reduce the similarity metric to a single value
requires 'cleanup';

sub sim_seq_search {

    my ( $self, $workload, %args ) = @_;

    croak 'expect an array(ref) as workload' unless ref $workload eq 'ARRAY';

    my $max_workers = $args{max_workers}
      // 1;    # set default value if not defined
    my $chunk_size = $args{chunk_size} // 1;  # set default value if not defined
    my $cleanup    = $args{cleanup}    // 1;  # set default value if not defined
    my @results;

    if ( $max_workers == 1 ) {
        foreach my $chunk ( @{$workload} ) {
            my $seq_align      = $self->seq_align->($chunk);
            my $sim_metric     = $self->extract_sim_metric->($seq_align);
            my $reduced_metric = $self->reduce_sim_metric->($sim_metric);
            push @results, $reduced_metric->@*; ## append to the results
            $self->cleanup( $seq_align, $sim_metric, $reduced_metric )
              if $cleanup;
        }
        return \@results;
    }
    else {
        my $mce = MCE->new(
            max_workers => $max_workers,
            chunk_size  => $chunk_size,
            gather      => MCE::Candy::out_iter_array( \@results ),
            user_func   => sub {
                my ( $mce, $chunk_ref, $chunk_id ) = @_;
                my @chunk_results;
                foreach my $chunk ( @{$chunk_ref} ) {
                    my $seq_align  = $self->seq_align->($chunk);
                    my $sim_metric = $self->extract_sim_metric->($seq_align);
                    my $reduced_metric =
                      $self->reduce_sim_metric->($sim_metric);
                    push @chunk_results, $reduced_metric->@*;
                    $self->cleanup( $seq_align, $sim_metric, $reduced_metric )
                      if $cleanup;
                }
                $mce->gather( $chunk_id, @chunk_results );
            }
        );
        $mce->process($workload);
    }
    return \@results;
}

1;
\end{lstlisting}
\newpage
\noindent A skeleton code that shows the use of the generic mapper and \textbf{LinearLinearGeneric} data flow is shown in \Fref{lst:SkeletonMapperCode}. The simple \textit{around} modifier, allows the user to develop the code for methods of a non-generic module without having to create a distribution, by providing access to \texttt{\$self}. Once the code has been fully fleshed out, then one simply copies the methods into the \texttt{pm} file of a distribution, and consumes the non-generic version of the data flow that is appropriate. At that stage, one can "downgrade" the object oriented module, i.e. from \texttt{Moose} to \texttt{Moo} if the full power of \texttt{Moose} would be a non-essential overkill for the task at hand.

\begin{lstlisting}[language=Perl,basicstyle=\footnotesize,frame=none,caption={Perl code for the \texttt{sim\_seq\_search} of the \textbf{SkeletonGenericMapper} role.},label={lst:SkeletonMapperCode},captionpos=b]
use MCE;
use Module::Find;
use Moose::Util qw( apply_all_roles );
use PDL::Lite;
use Bio::SeqAlignment::Components::SeqMapping::Dataflow::LinearLinearGeneric;
use Bio::SeqAlignment::Components::SeqMapping::Mapper::Generic;

my $mapper = Bio::SeqAlignment::Components::SeqMapping::Mapper::Generic->new(
    create_refDB       => \&create_db,
    use_refDB          => \&use_refDB,
    init_sim_search    => \&init_sim_search,
    seq_align          => \&seq_align,
    extract_sim_metric => \&extract_sim_metric,
    reduce_sim_metric  => \&reduce_sim_metric
);

$mapper->create_refDB->( $db_location, 'Hsapiens.cds.sample', \@dbfiles );
my $ref_DB = $mapper->use_refDB->( $db_location, 'Hsapiens.cds.sample' );
$mapper->sim_search_params( { '-m' => 'NW', '-n' => '0', '-l' => '' } ); 


apply_all_roles( $mapper,
    'Bio::SeqAlignment::Components::SeqMapping::Dataflow::LinearLinearGeneric'
);

my @workload = ... ;
$results        = $mapper->sim_seq_search( \@workload, max_workers => 4 );

sub init_sim_search {
    my ( $self, %params ) = @_;
    ...
}    

sub reduce_sim_metric {
    my ( $self, $sim_metric, %params ) = @_;
    ...
}

sub seq_align {
    my ( $self, $query_fname ) = @_;
    ...
}

sub extract_sim_metric {
    my ( $self,          $seq_align )           = @_;
    ...
}

sub create_db {
    my ( $self, $dbloc, $dbname, $files_aref ) = @_;
    ...
}

sub use_refDB {
    my $self = shift;
    ...
}
\end{lstlisting}
\newpage
\subsubsection{Enhancing \texttt{Edlib}'s performance with \texttt{MCE}, \texttt{FFI::Platypus}, \texttt{Inline::C} and \texttt{openMP}}
Having build a system that will allow us to "Jupyterize" code development for modules, we now use the mapper and data flow modules to enhance the performance of \texttt{Edlib}. The scripts for this session are found in the CPAN distribution \texttt{Bio::SeqAlignment::Examples::EnhancingEdlib} and in the author's main Github page\footnote{\url{https://github.com/chrisarg}}. To build these examples one needs to install an Alienized version of \texttt{Edlib},i.e. \texttt{Alien::SeqAlignment::edlib} that provides the command-line application, the static and the dynamic libraries of the aligner. For all the benchmarking experiments reported here, we used a sample of 1,648 sequences from the human transcriptome that were downloaded from Ensembl\footnote{\url{https://www.ensembl.org}}, the genome database project at the European Bioinformatics Institute. These sequences were used for as both queries and reference databases. When used as queries, these 1,648 sequences were split into two files of 1,000 and 648 sequences that had approximately the same size, i.e. about 800 kilobytes. In all examples, the similarity metric used was the edit distance and the reduction operation was simply to find the reference sequence with the minimum distance to the query. As the set of the query sequences was the same as the reference sequences, and we applied no noise to the queries, the minimum edit distance is zero. 
\paragraph{A parallel command-line \texttt{Edlib}} The command-line version of this aligner can only map multiple query sequences against a single reference sequence. To enhance this version, our database creation method \texttt{create\_refDB} had to split the reference sequences to multiple FASTA files (one for each reference sequence) and use the command line tool to map each of the two FASTA query files against these 1,648 files. In this case, the work elements that one may parallelize over is the either the query files or the reference sequence files. As we intended this example to be used as a test bed for wrapping other command-line wrappers that allow simultaneous mapping against multiple reference sequences, we parallelized over the query files. The \textbf{LinearLinearGeneric} data flow is applicable in the case of command-line mappers: the similarity metrics are computed separately for each sequence in the query files, which then have to be extracted from the output stream of the mapper and reduced. The reduction operation for the command line use case can be implemented in \texttt{PDL} as shown in \Fref{lst:ReductionCommandLine} from the script \texttt{testMap.pl} found in the Github and CPAN repositories for this section. the Mapping these 1,648 files against themselves with a single process in our machine took 254 seconds and with two processes 136 seconds.

\noindent\hspace{0.05\linewidth}\begin{minipage}{0.90\textwidth}
\begin{lstlisting}[language=Perl,basicstyle=\footnotesize,frame=none,caption={Parallel reduction of similarity metrics from the command-line \texttt{Edlib} using \texttt{PDL}.},label={lst:ReductionCommandLine},captionpos=b]
sub reduce_sim_metric {
    my ( $self, $sim_metric, %params ) = @_;
    my $db_in_use = $self->{refDB_access_params}->{DB};
    my $reduced_metric;
    while ( my ( $seq, $metric ) = each %{$sim_metric} ) {
        my $scores         = pdl($metric);
        my $best_score_idx = $scores->minimum_ind;
        my $best_match     = basename $db_in_use->{FILES}->[$best_score_idx];
        my $best_match_seqid =
          $db_in_use->{INDEX}{FNAME_SEQ_INDEX}{$best_match};
        push $reduced_metric->@*,
          [ $seq, $best_match_seqid, $scores->at($best_score_idx) ];
    }
    return $reduced_metric;
}
\end{lstlisting}
\end{minipage}

\paragraph{\texttt{Edlib\_MCE}: a parallel \texttt{Edlib} that uses \texttt{MCE}} For this enhancement we used the \texttt{FFI::Platypus} to interface with the dynamically linked library of \texttt{Edlib}. To do so, we wrote a module\footnote{\texttt{Bio::SeqAlignment::Components::Libraries::edlib}} to  access the API of \texttt{Edlib}. Our Perl API itself is rather simple, consisting of:
\begin{itemize}
\item three FFI records mapping to the C++ structures that \texttt{Edlib} utilizes to initialize a specific alignment task, provide character sets that should be considered equivalent for matching purposes and to return alignment, i.e. sequence similarity results.
\item five methods that create alignment configurations that instruct \texttt{edlib} which task to perform, manage the memory of alignment results (through \texttt{Edlib}) and execute the alignments. 
\end{itemize}
The data flow that is applicable in this case is the \textbf{LinearGeneric} one. The work elements in this case are the individual query sequences in a single FASTA file, and we parallelize by asking \texttt{MCE} through the \texttt{Bio::SeqAlignment::Components::SeqMapping::Dataflow::LinearGeneric} role to use one process per query sequence to map against the database of reference sequences. The logic of the \texttt{seq\_align} aligner/mapper method\footnote{This method does not handle yet the case of multiple "hits" of the reference in the query; this functionality will be implemented in a subsequent release. } is to map the query sequence by iterating over the reference ones in order to compute the alignments, collect the similarity metrics and finally reduce them using \texttt{PDL}. The component is written entirely in Perl, and interfaces with \texttt{Edlib} through the \texttt{FFI::Platypus} module. The code in \Fref{lst:MapperAlignerWO_OMP} does not allocate an intermediate Perl list to initialize a \texttt{PDL} object; rather it progressively extends a string that contains the similarity metrics separated by spaces, and utilizes \texttt{PDL}'s ability to initialize its container objects ("ndarrays") through space separated strings. Searching through the database of reference sequences is done through a hash that is created during the creation of the database of reference sequences\footnote{The hash itself is serialized by \texttt{Data::MessagePack}, and the resultant file is compressed by \texttt{Compress::LZF}.} and loaded when the method \texttt{use\_refDB } is called during the generic calling sequence shown in \Fref{lst:SkeletonMapperCode}.
 
\noindent\hspace{0.0\linewidth}\begin{minipage}{1\textwidth}
\begin{lstlisting}[language=Perl,basicstyle=\footnotesize,frame=none,caption={Aligner/mapper component for \texttt{Edlib}.},label={lst:MapperAlignerWO_OMP},captionpos=b]
sub seq_align {
    my ( $self, $query ) = @_;
    my $db_in_use       = $self->refDB_access_params->{DB};
    my $ref_seqs        = $db_in_use->{DATA};
    my $ref_seq_lengths = $db_in_use->{SEQLEN};
    my $query_length    = length $query->seq;
    my $query_seq       = $query->seq;
    my $edlib_config    = $self->refDB_access_params->{edlib_config};
    my $metric          = '';

    my $ref_seq_index = 0;
    for my $ref_seq (@$ref_seqs) {
        my $align =
          edlibAlign( $query_seq, $query_length,
            $ref_seq, $ref_seq_lengths->[$ref_seq_index],
            $edlib_config );
        $metric .= $align->editDistance . ' ';
        edlibFreeAlignResult($align);
        $ref_seq_index++;
    }

    my $scores         = pdl($metric);
    my $best_score_idx = $scores->minimum_ind;
    my $best_match_seqid =
      $db_in_use->{INDEX}{INDEX_SEQ_INDEX}{$best_score_idx};

    [ [ $query->id, $best_match_seqid, $scores->at($best_score_idx) ] ];
}
\end{lstlisting}
\end{minipage}

\paragraph{\texttt{Edlib\_MCE\_OpenMP}: a parallel \texttt{Edlib} that uses \texttt{MCE} and \texttt{OpenMP}}  While the command line version parallelized over FASTA files, and the library version accessed over \texttt{FFI::Platypus} parallelized over query sequences in files, the last parallel version of \texttt{Edlib}, parallelizes over both query and reference sequences using \texttt{Inline::C}. To do so, the \texttt{MCE} enabled \textbf{LinearGeneric} data flow is combined with (thread) level parallelism when aligning \textit{each} query sequence against the reference sequences using OpenMP. To leverage the capabilities of OpenMP for thread level parallelism a significant amount of the logic of the \texttt{seq\_align}, i.e. the loop over the reference sequences and the reduction operation had to be coded in C and interfaced with the \texttt{seq\_align} using \texttt{Inline::C}. This parallel version of \texttt{Edlib} \footnote{The relevant module is \texttt{Bio::SeqAlignment::Components::Libraries::edlib::OpenMP}}, is only possible because the OpenMP specification 5.0 and above allows one to use threads inside forked processes\footnote{The author found about this the very hard way, and named the relevant C function that allows the combination in the Perl application \texttt{\_fork\_around\_find\_out()}, i.e. FAFO.}. This additional enhancement comes at a substantial price, paid in lines of C code:
\begin{itemize}
\item While the the database of reference sequences is constructed in Perl using a hash structure as in the previous case, the  \textbf{actual addresses in DRAM} of the sequence strings must be made available to the C code to parallelize the search of a given query against the reference databases. To do so, we allocate an array to hold the addresses of the reference strings and their length in C and hand over its life cycle management to Perl using a macro \Fref{lst:SeqDBInC}. 
\item One needs ways to control the OpenMP environment, e.g. the assignment of workload to threads, i.e. the schedule and the number of threads, from within Perl. The module \texttt{OpenMP::Environment} allows many changes to be made to the OpenMP environment from within Perl, however the C code must also be able to read any changes made \textbf{after} the Perl application has been launched.  If any of the variables of the environment are not read from the C code, then they will not be set to the updated values. To make this point concrete, the function \texttt{\_ENV\_set\_num\_threads} in \Fref{lst:OMPAlignment} only updates the threads and not e.g. the scheduling of tasks by OpenMP.
\item Any data structures that were rather conveniently passed as FFI records, now must be passed using typemaps, or  as Scalar Values that hold an appropriately sized \textbf{non-null} terminated string. The latter is effectively a linear space in DRAM that holds the contents of the relevant C structure.
\item One has to allocate the memory to hold the mapping results in C, but hand over responsibility for management of this memory to Perl to avoid memory leaks and their security implications. 
\end{itemize}

\noindent\hspace{0.05\linewidth}\begin{minipage}{0.90\textwidth}
\begin{lstlisting}[language=C,basicstyle=\footnotesize,frame=none,caption={Pre-work to access Perl strings in C.},label={lst:SeqDBInC},captionpos=b]
// macro for the creation of a new SV buffer of a given size w/o resizing
#define NEW_SV_BUFFER(sv, buf , buffer_size) \
    SV *sv = newSV(0);                       \
    char *buf;                               \
    Newxz(buf,buffer_size,char);             \
    sv_usepvn_flags(sv, buf, buffer_size,SV_HAS_TRAILING_NUL|SV_SMAGIC)


// Define a struct to hold sequence data
typedef struct {
    uintptr_t seq_address;
    int seq_len;
} Seq;
SV* _make_C_index(AV* sequences) {
    int i;
    int n = av_len(sequences) + 1;
    _ENV_set_num_threads();
    size_t buffer_size = n * sizeof(Seq);      // how much space do we need?
    NEW_SV_BUFFER(retval, buf, buffer_size);
    Seq* RefDB = (Seq *)buf;
    #pragma omp parallel
    {
        int nthreads = omp_get_num_threads();
        size_t thread_id = omp_get_thread_num();
        size_t tbegin = thread_id * n / nthreads;
        size_t tend = (thread_id + 1) * n / nthreads;
        for (size_t i = tbegin; i < tend; i++) {
            SV** elem = av_fetch_simple(sequences, i, 0); // perl 5.36 and above
            STRLEN len = SvCUR(*elem);
            RefDB[i].seq_address = (uintptr_t)SvPVbyte_nolen(*elem);
            RefDB[i].seq_len = len;
        }
    }
    return retval;
}
\end{lstlisting}
\end{minipage}

\noindent\hspace{0.00\linewidth}\begin{minipage}{1\textwidth}
\begin{lstlisting}[language=C,basicstyle=\footnotesize,frame=none,caption={Query alignment, mapping and OpenMP environment reading in C},label={lst:OMPAlignment},captionpos=b]
// setting the openMP environment
void _ENV_set_num_threads() {
  char *num;
  num = getenv("OMP_NUM_THREADS");
  omp_set_num_threads(atoi(num));
}


// function that parallelizes the mapping of a single query seq over
// the database of reference sequences
AV *_edlib_align(char *query_seq, int query_len, SV *ref_DB, SV *config) {

  // initialization for the mapping
  int min_val = INT_MAX;
  int min_idx = -1;

  // get stuff from perl
  Seq *RefDB = (Seq *)SvPVbyte_nolen(ref_DB);
  int n_of_seqs = SvCUR(ref_DB) / sizeof(Seq); // size of database
  EdlibAlignConfig alignconfig = *(EdlibAlignConfig *)SvPVbyte_nolen(config);

  // prepare to send stuff back to perl
  AV *mapping = newAV();
  sv_2mortal((SV *)mapping);

  _ENV_set_num_threads();
#pragma omp parallel
  {
    int nthreads = omp_get_num_threads();
    size_t thread_id = omp_get_thread_num();
    size_t tbegin = thread_id * n_of_seqs / nthreads;
    size_t tend = (thread_id + 1) * n_of_seqs / nthreads;

    int thread_min_val = INT_MAX;
    int thread_min_idx = -1;
    for (size_t i = tbegin; i < tend; i++) {
      EdlibAlignResult align =
          edlibAlign(query_seq, query_len, (char *)RefDB[i].seq_address,
                     RefDB[i].seq_len, alignconfig);
      if (align.editDistance < thread_min_val) {
        thread_min_val = align.editDistance;
        thread_min_idx = i;
      }
      edlibFreeAlignResult(align);
    }
#pragma omp critical
    {
      unsigned int absthread_min_val = abs(thread_min_val);
      if (absthread_min_val < min_val) {
        min_val = absthread_min_val;
        min_idx = thread_min_idx;
      }
    }
  }
  // perl 5.36 and above
  av_push_simple(mapping, newSViv(min_idx)); // best match index
  av_push_simple(mapping, newSViv(min_val)); // match score (min edit distance)
  return mapping;
}

\end{lstlisting}
\end{minipage}
\\
The C function \texttt{\_edlib\_align} carries out the parallel sequence mapping by defining a parallel region through the compiler pragma \texttt{\#pragma omp parallel}. The code after the beginning of this block and until the critical region demarcated by the block \texttt{\#pragma omp critical} will be executed in parallel by the number of threads that were specified in the OpenMP environment variable \texttt{OMP\_NUM\_THREADS}. The critical region handles the reduction operation of finding the best scoring (smallest edit distance) reference sequence against the query processed by the function. While OpenMP offers explicit reduction clauses for simple functions (such as the minimum or the maximum), more complex ones have to follow fork-critical region synchronization design pattern shown here. 
\subsubsection{Performance benchmarking of the parallel versions of \texttt{Edlib} }
\begin{figure}[H]
\centering  
\includegraphics[scale=1]{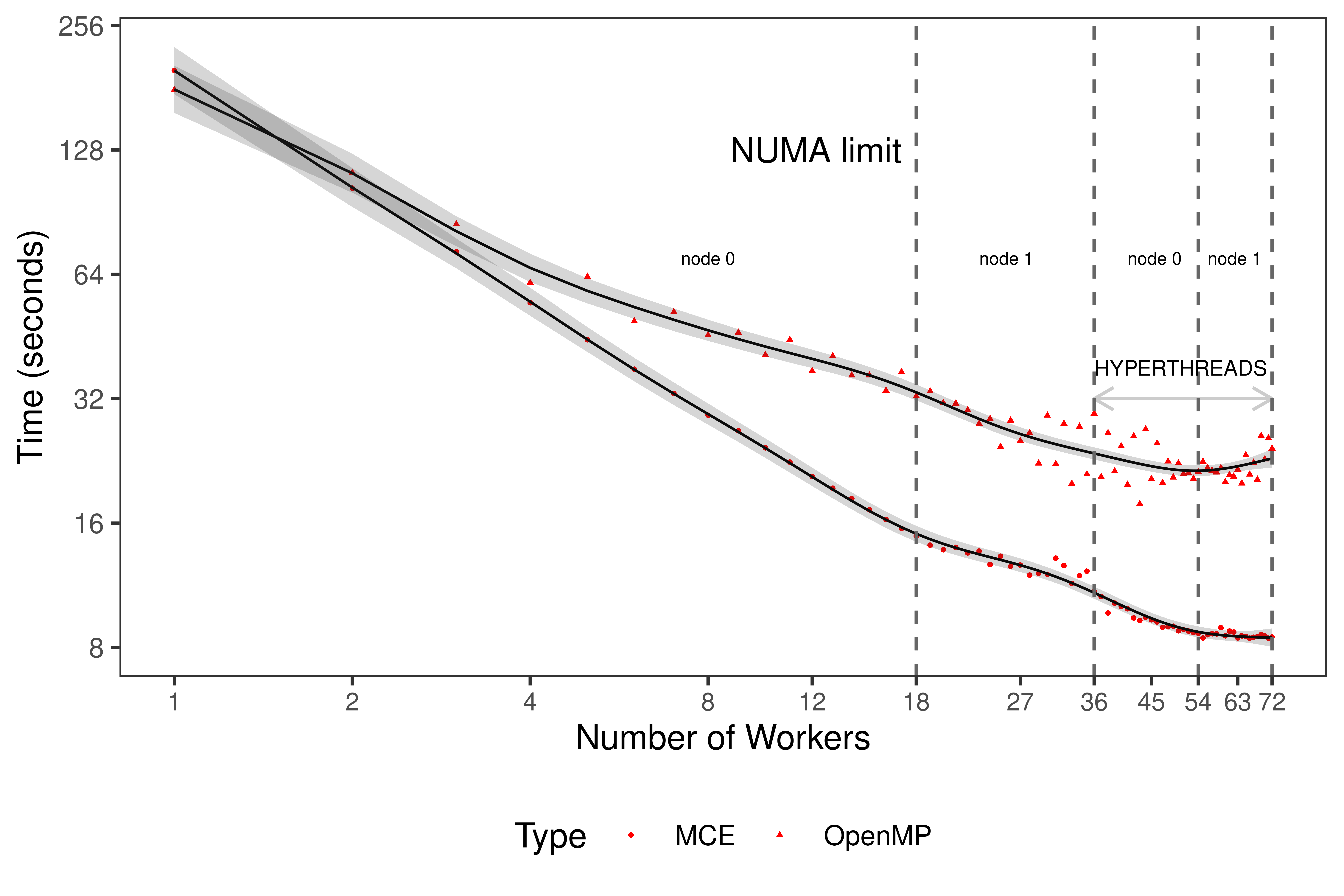}
\caption{Benchmarking a pure MCE and a pure OpenMP for an equal number of workers (processed under MCE or threads under OpenMP). The figure shows experimental data (OpenMP : triangles, MCE: filled circles), regression curve fits and 95\% confidence intervals (gray bands)}
\label{fig:EqNumOfWorkers}
\end{figure}

Having constructed \texttt{Edlib\_MCE} and \texttt{Edlib\_MCE\_OpenMP} that support process level and combined process and thread level parallelism, we now turn to the question : which one runs faster? In the first benchmark we varied the number of workers in \texttt{Edlib\_MCE} and compared them to the same number of OpenMP threads controlled by a single worker under \texttt{Edlib\_MCE\_OpenMP}, using the example dataset of 1,648 sequences discussed at the beginning of this section. The benchmarks are shown in \Fref{fig:EqNumOfWorkers}, which shows that process level parallelism scales much better than thread level parallelism for this application : there is a proportional decline in execution time until all the cores of the first node had been engaged, followed by a less steep, but still proportional decline as the threads in the second node of the Xeon processor are tasked. When the \texttt{MCE} starts assigning more processes than the physical cores, performance slows down and effectively stops by the time the hyperthreads of the second node are engaged. If the initial linear decline in execution time had held until all 72 cores had been engaged, regression analysis indicates that the task would have been completed in ~ 2 seconds (rather than 8). The OpenMP version demonstrates the same general pattern, i.e. a steep decline initially, that flattens later on, but tops out at a much higher execution time, likely because of poor memory locality.   

\begin{figure}[H]
\centering  
\includegraphics[scale=1]{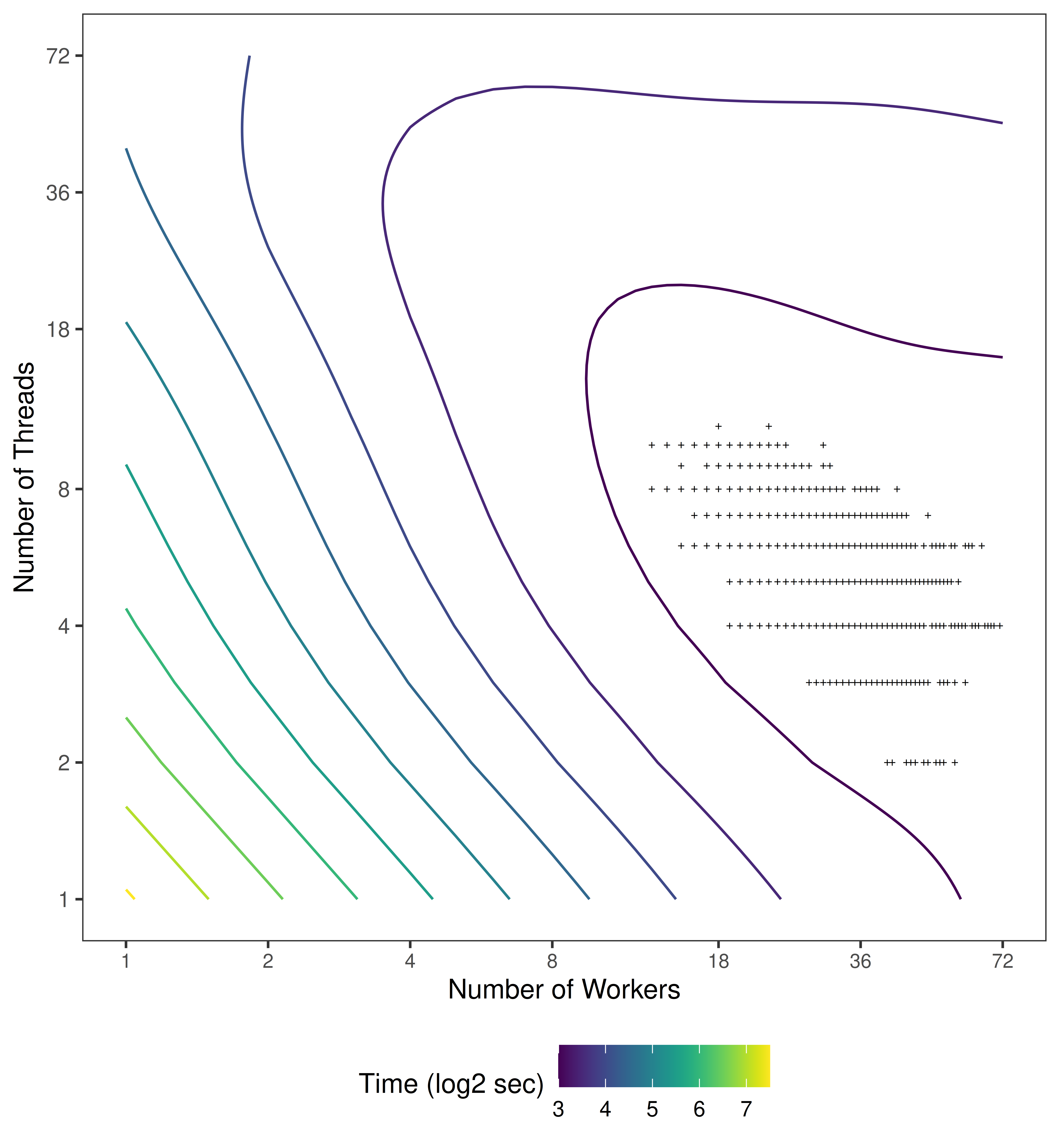}
\caption{Contour plot of the performance of \texttt{Edlib\_MCE\_OpenMP} over combinations of different number of threads and processes. The points highlighted via the '+' are the top 5\% performers(smallest execution time).  }
\label{fig:ProcessesThreadsContour}
\end{figure}

As the \texttt{Edlib\_MCE\_OpenMP} design allows us to profile variable combinations of processes and threads, we went on to examine all possible $72 \times 72 = 5,184$ possible combinations of processes and threads. These results are shown in \Fref{fig:ProcessesThreadsContour} as a contour plot of execution times. The best scoring combination used 25 processes and 4 threads per process to achieve a  predicted (via regression analysis) timing of 2.7 sec. These numbers  (as expected by the initial decline in execution time in \Fref{fig:EqNumOfWorkers}), effectively utilizing the full potential of the NUMA machine used for these experiments. 

\newpage
\section{Conclusions}
In this paper we examined the potential future role of Perl in the development of software components for data and compute intensive bioinformatic applications. In particular we used Perl to enhance the functionality of applications in R (\texttt{Polyester}) and C (\texttt{Edlib}). Both examples made heavy use of object oriented and meta-class programming techniques to abstract data flows in the enhanced applications, and in the latter case to "Jupyterize" the authoring of code for modules through the meta class capabilities in \texttt{Moose}. While seasoned Perl programmers have no issue in putting together modules at the blink of an eye, the current environment has a very strong preference for coding in a single file as much as possible. To provide such a  "Jupyter notebook" like coding experience for module codes in Perl, one has to provide some form of abstraction for the data flow and logic  as a scaffold for the code of the application. In the particular cases we considered, the data flows were simple enough for a single person to deliver, but in the general case the abstraction would require collaboration between software architects, engineers and end users for the abstraction to be created, before actual coding begins. Irrespective of the setting (solo coder v.s. a team), authoring the code for the abstraction is heavily aided by Perl's considerable ecosystem for MOP in the present and the ongoing work to bring OO inside the language through the \texttt{class} feature\footnote{\url{https://perldoc.perl.org/perlclass}}. We feel that Perl's OO and MOP capabilities endow the language with a considerable potential for application development in data and compute intensive fields such as Scientific Computing, Data Science and even Artificial Intelligence. In such fields, the compute intensive parts already exist as components in lower level language, yet the composition of these parts into fully fledged applications will require one to write the higher level components in some other language that the consumers will interact with. This is precisely the niche that was filled by R and increasingly by Python in these fields.  The \texttt{Edlib} enhancement shows that Perl can perform equally well in such tasks, while providing opportunities to combine process (at the data flow level) and thread (at the compute task level) parallelism to applications that will benefit from them. We illustrate that this is made possible by putting  OpenMP aware code in C under the control of the higher level abstractions in Perl and  used the \texttt{MCE} and \texttt{OpenMP::Environment} to experiment with the combination of thread and process parallelism that gave optimal performance. The particular example we considered, i.e. \texttt{Edlib} was rather extreme in that the library itself did not have multi-threaded capabilities to begin with. Hence a considerable amount of C code had to be written to provide this capability to it. More commonly, one will be dealing with multithreaded libraries, and thus the task will be considerably simpler, i.e. one would only have to interface with the external library through the \texttt{FFI::Platypus} interface, obviating the need to code in a foreign language. 
\\
At a more technical level, we found that all of our examples benefited from a judicious use of \texttt{PDL} for  vectorized, performant data intense operations. Vectorization is felt to be an important feature in scientific and data intense computing as it allows the programmer to transform and reduce large chunks of data with a single statement. While Perl built-in \texttt{map} and \texttt{grep} allow such operations to take place, users in data and science fields are accustomed to more direct ways to code them. Furthermore, they often expect such operations to occur \texttt{in-place}, i.e. without copying and moving large chunks of data in memory, something that is not easily obtained through \texttt{map}. In our examples we used \texttt{PDL} to vectorize random number generation in-place, achieving performance similar to that obtained in R, a language that comes out-of-the-box with vectorization capabilities. While this may seem as a niche application, it provides a self-contained case study for more complex use cases. Hence, we feel that the performance gains are generalizable; the performance implications are simply too large for Perl programmers to not consider \texttt{PDL} for data intense applications. Reduction operations are also aided greatly by using \texttt{PDL} in Perl code; while the specific reduction needs of our examples were rather simple by design, the large ecosystem of \texttt{PDL} statistical and numerical modules implies a considerably higher potential. For some applications \texttt{PDL} when combined with data flow and logic abstractions may allow one to avoid coding parts of the application in a foreign language. 
\\
In summary, we think that Perl should receive greater attention for the development of the high-level logic and the data flows in scientific and data intense applications. In the narrow field of bioinformatics in which the computations often map to discrete steps in experimental workflows, Perl can aid  authors by providing high level abstractions that allow them to work out the optimal combination of software components that can filter and process their voluminous data in a performant manner. The enhancements of \texttt{Edlib} we discussed in this paper highlight the path we are currently following in the \texttt{Bio::SeqAlignment} module that will deliver a set of components for novel applications for the analysis of third generation sequencing data.  

\newpage
\bibliography{PerlTalk}
\newpage
\begin{appendices}
\section{Hardware platform specifications}
All benchmarking took place with a Dual Xeon E2697v4 in an HP Z840 workstation. The NUMA architecture and the roofline diagram for the system are shown in the following figures.
\begin{figure}[H]
\renewcommand{\figurename}{Figure A}
  \centering
  \subfloat[][Topological map of the machine used for benchmarking]{\includegraphics[width=0.9\textwidth,height=0.40\textheight]{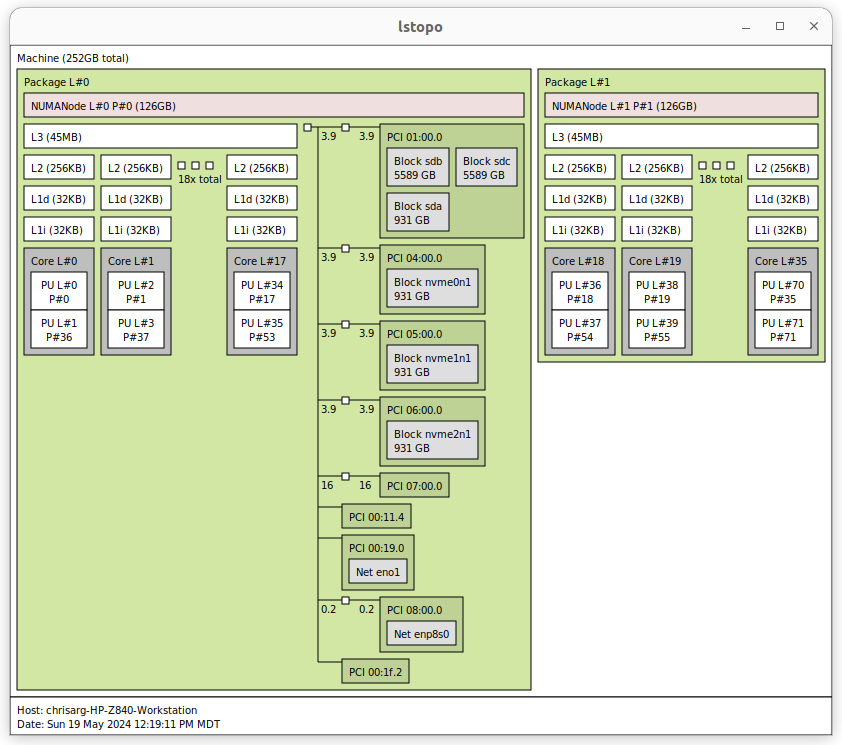} \label{fig:LSTOPO }} \\
  \subfloat[][Roofline diagram of the system used for benchmarking]{\includegraphics[width=0.9\textwidth,height=0.40\textheight]{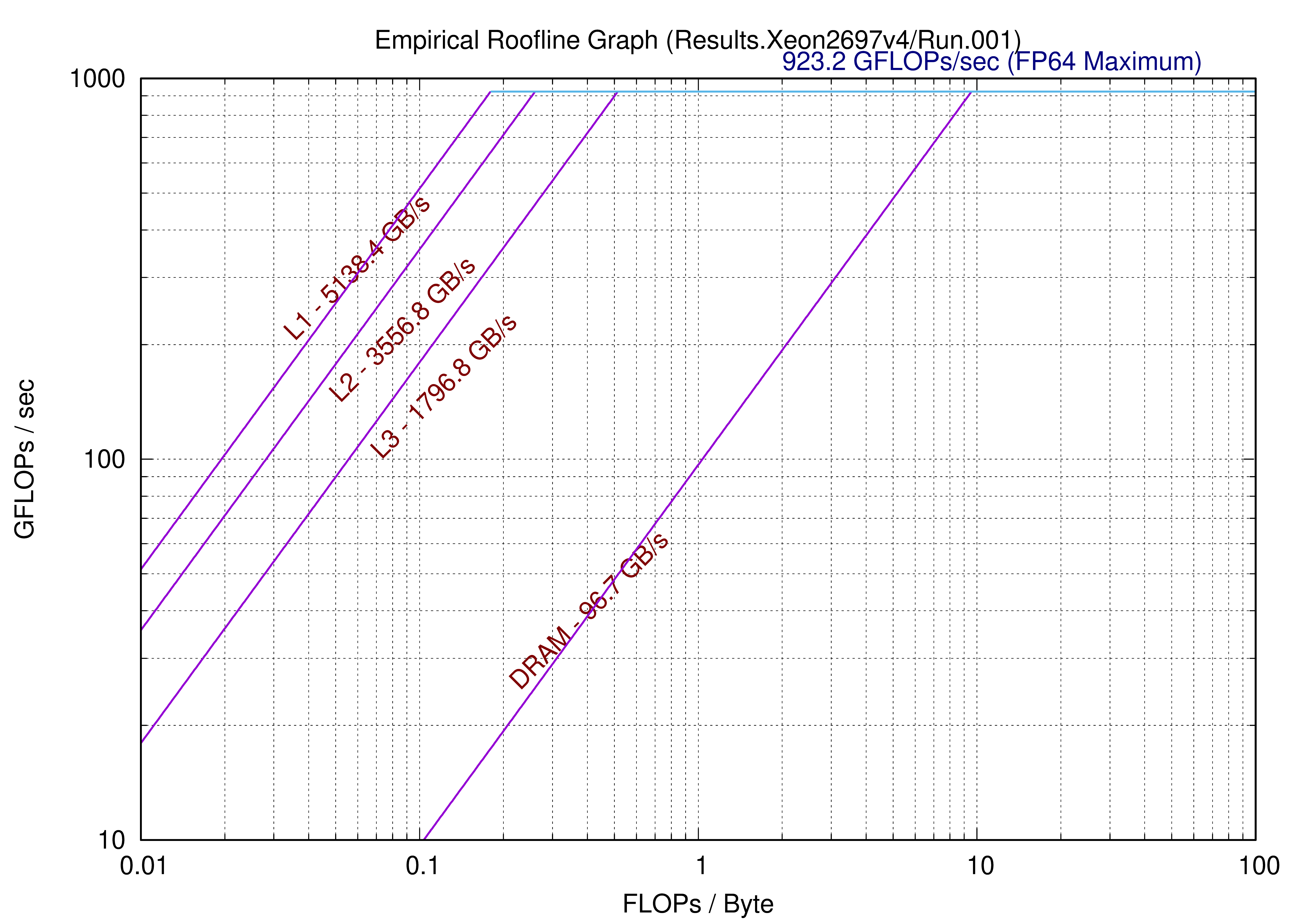} \label{fig:Roofline}}
\end{figure}
\newpage

\section{R code for truncated random number simulation}\label{apndx:RNGCodeInR} 
R code for the inverse CDF method for sampling from truncated distributions. The listing below provides code for sampling using eval strings, dereferencing of symbolic references and searching in the symbol table ("environment" in R) and finally fixing/hard-wiring the name of the dependencies e.g. the name of the RNG into the code. The R package \texttt{dqrng} was used to provide an interface to the Xoshiro RNG. The function \texttt{runif} is R's builtin RNG and is analogous to to Perl's \texttt{rand}.

\begin{lstlisting}[language=R,frame=single]
simtrunceval<-function(n,distr=NULL,uniformRNG=NULL,lower=NULL,
    upper=NULL,...){
  cdf<-function(x,...) eval(parse(text=paste("p",distr,sep="")))(x,...)
  invcdf<-function(x,...) eval(parse(text=paste("q",distr,sep="")))(x,...)
  uniform = function(x,...) eval(parse(text=uniformRNG))(x)
  lowercdf <-ifelse(is.null(lower),0,cdf(lower,...))
  uppercdf <-ifelse(is.null(upper),1,cdf(upper,...))
  domain <-uppercdf - lowercdf
  simvals<-uniform(n)
  simvals<-lowercdf + simvals*domain
  invcdf(simvals,...)
}

simtruncget<-function(n,distr=NULL,uniformRNG=NULL,lower=NULL,
    upper=NULL,...){
  cdf<-get(paste("p", distr, sep = ""))
  invcdf<-get(paste("q", distr, sep = ""))
  uniform = get(uniformRNG)
  lowercdf <-ifelse(is.null(lower),0,cdf(lower,...))
  uppercdf <-ifelse(is.null(upper),1,cdf(upper,...))
  domain <-uppercdf - lowercdf
  simvals<-uniform(n)
  simvals<-lowercdf + simvals*domain
  invcdf(simvals,...)
}

simtruncfixedrunif<-function(n,lower=NULL,upper=NULL,...){
  lowercdf <-ifelse(is.null(lower),0,plnorm(lower,...))
  uppercdf <-ifelse(is.null(upper),1,plnorm(upper,...))
  domain <-uppercdf - lowercdf
  simvals<-runif(n)
  simvals<-lowercdf + simvals*domain
  qlnorm(simvals,...)
}

simtruncfixedqrng<-function(n,lower=NULL,upper=NULL,...){
  lowercdf <-ifelse(is.null(lower),0,plnorm(lower,...))
  uppercdf <-ifelse(is.null(upper),1,plnorm(upper,...))
  domain <-uppercdf - lowercdf
  simvals<-dqrunif(n)
  simvals<-lowercdf + simvals*domain
  qlnorm(simvals,...)
}
\end{lstlisting}
\end{appendices}
\newpage

\end{document}